\documentclass[reqno,12pt]{article}
\usepackage{amsmath,amsfonts,amssymb,amsthm,amstext,amscd,eucal,xcolor}
\usepackage[all]{xy}
\usepackage{amsmath, amssymb}
\usepackage{mathtools}
\newcommand{\mathsym}[1]{{}} 
\usepackage[colorlinks=true,
            linkcolor=blue,
            urlcolor=blue,
            citecolor=blue]{hyperref}
\usepackage{enumerate}

\usepackage{graphicx}% Include figure files
\usepackage{amsmath} \usepackage{amssymb}
\usepackage{bm}% bold math
\usepackage{cite,hyperref}
\usepackage{mathtools}
\usepackage{amsmath} \usepackage{amssymb}
\usepackage{bm}% bold math
\usepackage{graphicx}
\usepackage[active]{srcltx}
\makeatletter \@addtoreset{equation}{section}

\makeatletter\renewcommand\section{\@startsection {section}{1}{\z@}%
                                   {-3.5ex \@plus -1ex \@minus -.2ex}%nn
                                   {2.3ex \@plus.2ex}%
                                   {\normalfont\large\bfseries}}
\renewcommand\subsection{\@startsection{subsection}{2}{\z@}%
                                     {-3.25ex\@plus -1ex \@minus -.2ex}%
                                     {1.5ex \@plus .2ex}%
                                     {\normalfont\bfseries}}

\newcommand{\be}{\begin{equation}}
\newcommand{\ee}{\end{equation}}
\newcommand{\bea}{\begin{eqnarray}}
\newcommand{\eea}{\end{eqnarray}}
\newcommand{\bse}{\begin{subequations}}
\newcommand{\ese}{\end{subequations}}
\newcommand{\beqa}{\begin{eqnarray}}
\newcommand{\eeqa}{\end{eqnarray}}
\newcommand{\beqar}{\begin{eqnarray*}}
\newcommand{\eeqar}{\end{eqnarray*}}
\newcommand{\bi}{\begin{itemize}}
\newcommand{\ei}{\end{itemize}}
\newcommand{\bn}{\begin{enumerate}}
\newcommand{\en}{\end{enumerate}}
\newcommand{\ba}{\begin{array}}
\newcommand{\ea}{\end{array}}
\newcommand{\bc}{\begin{center}}
\newcommand{\ec}{\end{center}}

\definecolor{darkgreen}{rgb}{0,0.3,0}
\definecolor{darkblue}{rgb}{0,0,0.3}
\definecolor{darkred}{rgb}{0.7,0,0}

\makeatother
\begin{document}

\newcommand{\email}[1]{\footnote{E-mail: \href{mailto:#1}{#1}}}

\title{\bf\Large{Perturbative effective action for the photon in
noncommutative QED$_{2}$ and exactness of the Schwinger mass}}

\author{\bf{M.~Ghasemkhani} \email{ghasemkhani@ipm.ir } $^{a}$, \bf{A.~A.~Varshovi} \email{ab.varshovi@sci.ui.ac.ir} $^{b,c}$ and \bf{R.~Bufalo} \email{rodrigo.bufalo@dfi.ufla.br} $^{d}$ \\\\
%\vspace{1cm}
%\normalsize
\textit{\small$^a$  Department of Physics, Shahid Beheshti University,  G.C., Evin, Tehran 19839, Iran}\\
\textit{\small $^b$   Department of Mathematics, University of Isfahan,
Isfahan 81745-163, Iran}\\
 \textit{\small $^c$   School of Mathematics, Institute for Research in Fundamental
Sciences (IPM), }\\
\textit{\small P.O. Box 19395-19395-5746., Tehran, Iran}\\
\textit{\small $^d$   Departamento de F\'isica, Universidade Federal de Lavras,}\\
\textit{\small Caixa Postal 3037, 37200-000 Lavras, MG, Brazil}\\
}
%%%%%%%%%%%%%%%%%%%%%%%%%%%%%%%%%%%%%%%%%%%%%%%%%%%%%%%%%%%%%%%%%%%%%%%%%%%%%%%%%%%%%%%%
%\date{\today}
\maketitle

\begin{abstract}
In this paper, we  discuss the noncommutative QED$_2$ in the S-matrix framework.
We are interested in perturbatively proving that the  exact Schwinger
mass $\mu^2=\frac{e^2}{\pi}$
 does not receive noncommutative corrections to any order in loop
expansion. In this sense,
the S-matrix approach is useful since it allows us to work with the
effective action $\Gamma[A]$
(interaction term) to compute the corresponding  gauge field 1PI
two-point function
at higher orders. Furthermore, by means of $\alpha^*$-cohomology, we generalize
the QED$_2$
S-matrix analysis in the Moyal star product to all translation-invariant
star products.
\end{abstract}

%\vspace{0.5in}

%%%%%%%%%%%%%%%%%%%%%%%%%%%%%%%%%%%%%%%%%%%%%%%%%%%%%%%%%%%%%%%%%%%%%%%%%%%%%%%%%%%%%%%%%%
\setcounter{footnote}{0}
\renewcommand{\baselinestretch}{1.05}  %Line spacing
%%%%%%%%%%%%%%%%%%%%%%%%%%%%%%%%%%%%%%%%%%%%%%%%%%%%%%%%%%%%%%%%%%%%%%%%%%%%%%%%

%\addtocontents{toc}{\protect\setcounter{tocdepth}{2}}
\newpage
\tableofcontents
%%%%%%%%%%%%%%%%%%%%%%%%%%%%%%%%%%%%%%%%%%%%%%%%%%%%%%%%%%%%%%%%%%%%%%%%%%%%%%%%%%%%%%%%%%%
\section{Introduction}
\label{sec-intro}

  Because of the lack of proper answers to some important questions in our description of nature, it is rather natural to look for alternative proposals in order to gain insights to improve our knowledge and fundamental theories. In this sense, reducing the spacetime dimensionality has proved to be one of the richest scenarios where ideas can be successfully tested.
Low-dimensional field theory models are widely used in the description of many physical phenomena, whose applicability range from planar physics in condensed matter $(2+1)$ spacetime \cite{Jackiw:1989nq} to many fermions and integrable systems in $(1+1)$ spacetime \cite{stone}.
In particular, despite sharing relevant phenomena with more realistic models, many two-dimensional field theory models have a distinguishable property of being exactly solvable such as fermionic quartic interactions (Thirring model) \cite{Thirring:1958in} and quantum electrodynamics (QED$_2$) \cite{schwinger-1,schwinger-2}.

The most interesting two-dimensional field theory model is the massless QED$_2$, known as the Schwinger model. This model displays two important
features, that local gauge symmetry does not necessarily imply the existence of a massless gauge field, which is an example of dynamical mass generation, and that, due to the linear behavior of the electrostatic potential, fermions do not appear in the physical states and hence the spectrum of this theory includes a free massive boson. In summary, this shows that the fermionic field is confined, which is in total analogy with the quark confinement phenomenon happening in quantum chromodynamics (QCD) -- in this sense massless QED$_2$ is considered as a toy model for QCD$_4$.
All of these interesting aspects have been extensively studied in several works
such as \cite{schwinger-1,schwinger-2,casher,swieca,abdalla}.

On the other hand, in the pursuit of a better understanding of quantum behavior of gravity we have witnessed an intense activity in the last few decades in attempts to unveil the structure of nature at very short distances \cite{AmelinoCamelia:2008qg}, from which the most important theories we might refer to are string theory and quantum loop gravity, etc. Moreover, among the several interesting features resulting from such descriptions, the quantum realm of systems under the influence of strong gravitational fields, one can say that the loss of the smoothness of the spacetime and that a minimal length is necessarily present are the most striking ones \cite{Hossenfelder:2012jw}. A framework which  encompasses all the desired features is the noncommutative spacetime, where the spacetime coordinates are noncommutative, satisfying an algebra $\left[\hat{x}_\mu , \hat{x}_\nu\right]=i\Theta_{\mu\nu}$, where $\Theta_{\mu\nu}$ is a skew-symmetric matrix, measuring the spacetime noncommutativity through uncertainty relation $\Delta \hat{x}_\mu \Delta \hat{x}_\nu \sim \left|\Theta_{\mu\nu}\right|$ \cite{witten}.

Interestingly, spacetime noncommutativity gives rise to new and interesting features on field theories, basically inducing a non-Abelian structure on the fields due to the star product. Such features have been extensively studied in many different field theoretical contexts; some recent studies are Refs.~\cite{witten,gomis,jabbari,Chaichian:2001py,Blaschke:2009rb,DAscanio:2016jmt,steinacker,arzano,ghasemkhani-16}.
Since the Schwinger model has played an important part as a QCD$_4$ toy model,
naturally it was analyzed in many contexts by the introduction of spacetime noncommutativity, unveiling interesting properties of noncommutative QED$_2$ \cite{grosse,saha,rahaman,ghasemkhani-9,ghasemkhani-11,Armoni:2011pa,petrov,ghasemkhani-13}.

Among the new features analyzed in the noncommutative Schwinger, one may see that the noncommutativity does not affect the deconfining behavior of the model \cite{rahaman}. Now with regard to the dynamical mass generation for the gauge field, it has been shown that the Schwinger mass does not receive any noncommutative corrections at one-loop level and from higher orders as well\cite{ghasemkhani-11,ghasemkhani-13}. Furthermore, one-loop analysis for the noncommutative supersymmetric Schwinger model indicates that the generated mass for the gauge superfield is independent from the noncommutative
parameter \cite{petrov}.

Although many features of QED$_2$ have been extensively studied in a noncommutative framework, we believe that there are some blank points which deserve more attention, mainly concerning correction to the photon mass. Our interest here might be seen as a continuation of \cite{ghasemkhani-13} in which the contribution of the higher-order graphs to the photon self-energy using the standard method was computed. While in the present work, we propose the S-matrix approach for the photon one-loop effective action $\Gamma[A]$
to more conveniently concentrate on the analysis of the higher-loop noncommutative corrections to the physical pole of the photon propagator.
Actually, in this approach, the S-matrix analysis is applied to $\Gamma[A]$ and then diagrams with a single (multi)-fermion loop(s) at any order are generated. Furthermore, as a bonus of our analysis, we show that the calculation of the relevant noncommutative phase factor for any complicated diagram can be exactly specified in an elegant way, without resorting to the Feynman rules used in \cite{ghasemkhani-13}.

In this paper, in order to examine higher-order contributions to the photon 1PI two-point function, we will approach noncommutative Schwinger model in a S-matrix framework.
We start Sec.~\ref{sec1} by establishing the main aspects of the photon one-loop effective action $\Gamma[A]$, as well presenting some statements involving this functional in commutative and noncommutative spacetime.
In Sec.~\ref{sec2}, we write down and compute explicitly high-order elements of the S-matrix for the effective action $\Gamma[A]$. This analysis provides us with powerful but easy tool to determine the respective contributions order by order to the 1PI photon 2-point function.
Afterwards, in Sec.~\ref{sec4} we generalize the obtained results in the Moyal star product to all translation-invariant star products by means of $\alpha^*$-cohomology.
In Sec.~\ref{sec5} we summarize the results and present our final remarks.

%%%%%%%%%%%%%%%%%%%%%%%%
\section{One-loop effective action}
\label{sec1}
%%%%%%%%%%%%%%%%%%%%%%%%%%%%%%%%%%%%%%%%%%%%%%%%%%%%%%%%%%%
In the present section, we introduce the noncommutative Schwinger model by constructing the noncommutative extension of massless fermionic fields interacting
 with an Abelian gauge field. For this purpose, we consider the following action
\begin{equation}
{\cal{S}}=\int d^{2}x\left[\bar\psi\star i\gamma^{\mu}D_{\mu}\psi-\frac{1}{4}F_{\mu\nu}\star F^{\mu\nu}+{\cal{L}}_{g.f}+{\cal{L}}_{gh}\right],
\label{eq:a1}
\end{equation}
 where the covariant derivative is defined as $D_{\mu}\psi=\partial _{\mu} \psi+ieA_{\mu}\star\psi$ and the field strength tensor $F_{\mu\nu}=\partial_{\mu}A_{\nu}-\partial_{\nu}A_{\mu}+ie[A_{\mu},A_{\nu}]_{\star}$, where $[~ ,~ ]_{\star}$ is the Moyal bracket. The Moyal star product between the functions $f(x)$ and $g(x)$ is defined as
\begin{equation}
f\left(x\right)\star g\left(x\right)=f\left(x\right)\exp\Big(\frac{i}{2}
\Theta ^{\mu\nu}\overleftarrow{\partial_{\mu}}~\overrightarrow{\partial_{\nu}}\Big)g\left(x\right).
\label{eq:a3}
\end{equation}
The gauge fixing and the ghost terms in the action \eqref{eq:a1} can be written in a BRST exact form \cite{weinberg}
 \begin{align}
 {\cal{L}}_{g.f}+ {\cal{L}}_{gh} = s \Big(\bar{c} \star \Big[\frac{\xi}{2}B -\Omega[A]\Big] \Big)
 \label{eq:a2}
 \end{align}
where $B$ is the Nakanishi-Lautrup field and $\Omega$ is a generic gauge condition, i.e., $\Omega [A]=0$. The above Lagrangian \eqref{eq:a1} is invariant under the BRST Slavnov transformations
\begin{align}
sA_\mu =D_\mu c, \quad s\psi =ie c\star \psi , \quad sc=ie c\star c, \quad s\bar{c}= -B, \quad s B=0.
\end{align}

In what follows we shall take the following considerations: the noncommutative structure of the spacetime is determined as usual by $[x^\mu,x^\nu]=i\Theta ^{\mu\nu}$, in which $\Theta^{\mu\nu}$ is a constant antisymmetric parameter, containing one nonvanishing component $\Theta^{01}$ in two dimensions.
However, in order to avoid the unitarity violation, we choose to work in Euclidean signature \cite{gomis}. Furthermore, for the sake of simplicity, the light-cone coordinates $x_{\pm}=x_{1}\pm ix_{2}$ are chosen.
In a two-dimensional gauge theory, the light-cone gauge $A_{-}=0$ is a rather suitable condition, since, in this gauge, the photon self-interaction terms are eliminated and the gauge theory is ghost free, similarly written as the commutative Abelian gauge theories \cite{burnel}.

Now, in order to analyze the one-loop effective action for the photon, we shall use the Feynman path-integral formalism. The path integral of the present model is readily given by
\begin{align}
{\cal{Z}}=\int {\cal{D}}A_{\mu}{\cal{D}}\bar\psi {\cal{D}}\psi~e^{i\left({\cal{S}}_{g}+{\cal{S}}_{f}\right)}.
\label{eq:a4}
\end{align}
 Here, we have separated the total action ${\cal{S}}$ \eqref{eq:a1} into two parts ${\cal{S}}_{g}$ and ${\cal{S}}_{f}$ indicating the pure gauge and the fermionic matter parts, respectively. The action ${\cal{S}}_{g}$ in the light-cone gauge includes the third and the fourth terms of the action \eqref{eq:a1}, without a ghost term.
Integrating out the fermionic fields yields the one-loop effective action for the gauge field
\begin{equation}
e^{i\Gamma[A]}=\int {\cal{D}}\bar\psi {\cal{D}}\psi~e^{i{\cal{S}}_{f}},
 \label{eq:a5}
\end{equation}
where the effective action $\Gamma[A]$ is defined as follows
\begin{equation}
\Gamma\left[A\right]=\ln\frac{\det\left(i\displaystyle{\not}\partial-e\displaystyle{\not}A\star\right)}
{\det \left(i\displaystyle{\not}\partial\right)}
=-\sum_{n=1}^{\infty}\frac{1}{n}tr\Big[\left(\frac{i}{i\displaystyle{\not}\partial}\right)(-ie \displaystyle{\not}A\star)\Big]^{n}.
\label{eq:a6}
\end{equation}

\begin{figure}[t]
\vspace{-1.4cm}
\centering
  \includegraphics[width=4cm,height=3cm]{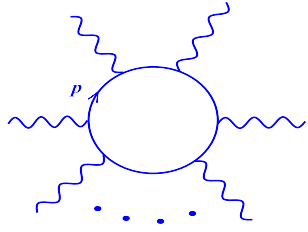}
      \caption{The relevant graph for the $n$ th term of the one-loop effective
  action.
}\label{generic-term}
\end{figure}

Because of the perturbative structure of noncommutative QED, the obtained series is asymptotically convergent and thus the partition function \eqref{eq:a4} is revised as
\begin{equation}
{\cal{Z}}=\int {\cal{D}}A_{\mu}~e^{i\left({\cal{S}}_{g}+\Gamma[A]\right)}.
\label{eq:a-7}
\end{equation}
Since our main focus is the perturbative calculation of $\Gamma[A]$, it is more convenient to replace Eq.~\eqref{eq:a6} by the following expression
\begin{equation}
\Gamma\left[A\right]=\sum_{n=1}^{\infty}\int\prod_{i=1}^{n} d^{2}x_{i} ~A_{\mu_{1}}\left(x_{1}\right)\ldots A_{\mu_{n}}\left(x_{n}\right)\Gamma^{\mu_{1}\ldots\mu_{n}}\left(x_{1},\ldots,x_{n}\right),
\label{eq:a8}
\end{equation}
where $\Gamma^{\mu_{1}\ldots\mu_{n}}$ refers to a one-fermion-loop graph with $n$ external photon lines, as shown in Fig.~\ref{generic-term}. The explicit form of the tensor $\Gamma^{\mu_{1}\ldots\mu_{n}}$ is illustrated as
\begin{align}
\Gamma^{\mu_{1}\ldots\mu_{n}}\left(x_{1},\ldots, x_{n}\right)&=-\frac{\left(-e\right)^{n}}{n}\int\prod_{i=1}^{n}\frac{d^{2}k_{i}}{\left(2\pi\right)^{2}}
\left(2\pi\right)^{2}\delta(\sum_{i}k_{i})~e^{-i\sum\limits_{i}k_{i}.x_{i}} e^{-\frac{i}{2}\sum\limits_{i<j}k_{i}\wedge k_{j}}e^{p\wedge \sum\limits_{i=1}^{n}k_{i}}\nonumber\\
 &\times\Xi^{\mu_{1}\ldots\mu_{n}}\left(k_{1},\ldots,k_{n-1}\right),
 \label{eq:a-9}
\end{align}
 with the notation $k_{i}\wedge k_{j}=\Theta^{\mu\nu}k_{i\mu}k_{j\nu}$ and the expression $\Xi^{\mu_{1}\ldots\mu_{n}}$ is given by
\begin{align}
\Xi^{\mu_{1}\ldots\mu_{n}}=
\int\frac{d^{2}p}{\left(2\pi\right)^{2}}
\frac{tr\Big[\gamma^{\mu_{1}}\left(\displaystyle{\not}p+\displaystyle{\not}k_{1}
\right)\gamma^{\mu_{2}}
\left(\displaystyle{\not}p+\displaystyle{\not}k_{1}+\displaystyle{\not}k_{2}\right)\gamma^{\mu_{3}}
\ldots (\displaystyle{\not}p+\sum\limits_{i=1}^{n-1}\displaystyle{\not}k_{i})
\gamma^{\mu_{n}}(\displaystyle{\not}p)\Big]}
{
p^{2}\left(p+k_{1}\right)^{2} \left(p+k_{1}+k_{2}\right)^{2}
\left(p+\sum\limits_{i=1}^{n-1}k_{i}\right)^{2}}
,\label{eq:a10}
\end{align}
where $p$ is the momentum of fermionic loop and $k_{i}$'s are the momenta of the external gauge fields.

%%%%%%%%%%%%%%%%%%%%%%%%%%%%%%
%%%%%%%%%%%%%%%%%%%%%%%%%%%%%%

As is easily seen in Eq.~\eqref{eq:a10}, the effect of the noncommutativity has been encoded into two phases: $e^{-\frac{i}{2}\sum\limits_{i<j}k_{i}\wedge k_{j}}$ and $e^{p\wedge \sum\limits_{i=1}^{n}k_{i}}$, although the second one is removed due to the energy-momentum conservation constraint. Therefore, the noncommutative phase factor does not depend on the momentum of the fermion loop and this property is independent from the number of external legs.

Before proceeding with our analysis, it is worth presenting some overall comments about the general structure of $\Gamma[A]$, in particular, its initial terms in the series form \eqref{eq:a6}, for both commutative and noncommutative QED in various dimensions:
 %%%%%%%%%%%%%%%%%%%%%%%%%%%%%%%%%%%%%%%%%%
 \begin{enumerate}

 \item  \textbf{Commutative case}
 %%%%%%%%%%%%%%%%%%%%%%%%%%%%%%%%%%%%%%%%%%
 \begin{description}
   \item[$~~~~\bullet~$\emph{d=4}:]~The first term of \eqref{eq:a6} contributing to the photon one-point function vanishes whereas the second term gives the nonzero one-loop correction to the two-point gauge field function, without any mass generation for the photon. The next term, $n=3$, includes the contribution of two triangle diagrams with two different orientations of the fermion loop which cancel mutually, according to the Furry's theorem: diagrams with an odd number of external photons are vanishing.

Moreover, for $n=4$, there are six independent diagrams whose divergences cancel each other and we are left with a finite result producing a four-photon interaction term named as the Euler-Heisenberg effective action for the soft photons (with lower energy than the fermion mass) \cite{heisenberg,dittrich}. In the case of other remaining terms, i.e., $n=2k$ with $k\geq 3$, we expect to have graphs with finite amplitudes since QED is a renormalizable quantum field theory in four dimensions.
%%%%%%%%%%%%%%%%%%%%%%%%%%%%%%%%%%%%%%%
   \item[$~~~~\bullet~$\emph{d=3}:]~We first notice that QED$_{3}$ is a super-renomalizable theory, giving us a finite value for the integrals appearing at loop level. Second, since charge conjugation is a symmetry of QED Lagrangian in any dimensions, then QED$_{3}$ is charge conjugation invariant as well. This invariance allows us to use Furry's theorem to conclude that the amplitudes of graphs with odd values of $n$ are readily removed.

Besides, by taking into account these considerations, we realize that the nonzero terms of the series expansion \eqref{eq:a6} are given by $n=2k$ with $k\geq 1$ which produce the finite amplitudes.

For $n=2$, a finite amplitude for the vacuum polarization tensor is obtained, consisting of a even and odd parity parts, whereas in $d=4$ a single even parity divergent amplitude is found. The presence of this additional odd-parity term originates from the trace of an odd number of gamma matrices, which in three dimensions leads to topological mass generation for the photon. In configuration space, these even and odd-parity parts induce the ordinary Maxwell-Chern-Simons action, at the large fermion mass limit \cite{redlich}. In the next to leading order of $n=2$, gauge invariant higher-derivative contributions to the Maxwell-Chern-Simons action are found \cite{jackiw}.
%%%%%%%%%%%%%%%%%%%%%%%%%%%%%%%%%%%%%%%
   \item[$~~~~\bullet~$\emph{d=2}:]~In this dimension, we have a super-renormalizable theory which is analytically
solvable for massless fermions, named as the Schwinger model, while the massive Schwinger model does not have an exact solution \cite{schwinger-2,casher}. Once again, due to the Furry's theorem, we are left with only even terms, $n=2k$ in \eqref{eq:a6}, which for $n=2$, we observe a dynamical mass generation for the photon at one-loop order. However, the nonperturbative analysis demonstrates that $\Gamma[A]$ for the Schwinger model is exactly
determined \cite{abdalla,sorensen}, presented as
\begin{equation}
\Gamma[A]=\frac{e^{2}}{\pi}\int\frac{d^{2}k}{(2\pi)^{2}}\widetilde{A}_{\mu}(k)\Big(g^{\mu\nu}-
\frac{k^{\mu}k^{\nu}}{k^{2}}\Big)\widetilde{A}_{\nu}(-k),
\label{eq:a11}
\end{equation}
which explicitly shows that the photon has received a mass term, Schwinger mass, $\mu^{2}=\frac{e^{2}}{\pi}$; moreover, this mass generation is compatible with gauge invariance \cite{schwinger-2}. Indeed, by comparing Eqs.~\eqref{eq:a11} and
\eqref{eq:a8}, it is immediately seen that $\Gamma^{\mu_{1}\ldots\mu_{n}}$ is nonzero only for $n=2$. This means that there is only a quadratic term for the gauge field in the effective action $\Gamma[A]$, while the remaining terms are vanishing, which is in contrast to the four-dimensional case. The perturbative analysis indicates that the Schwinger mass is one-loop exact and does not receive any additional corrections from higher loops.  A detailed perturbative proof on the exactness of the Schwinger mass is provided in appendix \ref{sec-appA}.
 \end{description}

%%%%%%%%%%%%%%%%%%%%%%%%%%%%%%%%%%%%%%%%%%
 \item  \textbf{Noncommutative case}
%%%%%%%%%%%%%%%%%%%%%%%%%%%%%%%%%%%%%%%%%%
\begin{description}
  \item[$~~~~\bullet~$\emph{d=4}:]~The term $n=1$ does also vanish and the amplitude of the second term $n=2$ is the same as its commutative counterpart, a divergent quantity with even parity. Unlike the commutative case, the next term $n=3$, corresponding to the one-loop correction to the cubic gauge vertex, is not zero since Furry's theorem is not applicable here \cite{jabbari}. Finally, for $n=4$, we obtain a one-loop correction to the quartic gauge vertex which is divergent as well, in contrast to the finite result arising from commutative analysis; further details are found in Ref.~\cite{charneski}.
%%%%%%%%%%%%%%%%%%%%%%%%%%%%%%%%%%%%%%%%%%
  \item[$~~~~\bullet~$\emph{d=3}:]~The absence of charge conjugation symmetry in the noncommutative setup imposes that $n=2k+1$ graphs are now nonvanishing. The explicit computations of the terms $n=2,3,4$ lead to three finite amplitudes, in which the sum of them induces the NC-Maxwell-Chern-Simons theory, at the large fermion mass
      limit \cite{chu,banerjee}.
              In the next to leading order, the higher-derivative contributions to the NC-Maxwell-Chern-Simons action, together with a discussion on the gauge invariance of the induced higher-derivative terms, have also been investigated in \cite{ghasemkhani-14}.
%%%%%%%%%%%%%%%%%%%%%%%%%%%%%%%%%%%%%%%%%
  \item[$~~~~\bullet~$\emph{d=2}:]~There are several papers on the study of noncommutative extension of
  the Schwinger model and its bosonized version, e.g., Refs.~\cite{grosse,saha,ghasemkhani-11,ghasemkhani-13}, where in \cite{ghasemkhani-11,ghasemkhani-13}, it is shown that the Schwinger mass, $\mu^{2}=\frac{e^{2}}{\pi}$, remains unchanged and does not receive any noncommutative corrections.
\end{description}
 \end{enumerate}

In what follows, we will illustrate how we can generate the higher-order contributions to the photon $n$-point functions using the effective action $\Gamma[A]$, Eq.~\eqref{eq:a8}. We are interested, in particular, in computing whose graphs which give higher-loop corrections to the photon self-energy in the presence of massless fermions. To this end, we shall proceed in computing the S-matrix elements for the effective action $\Gamma[A]$.

%%%%%%%%%%%%%%%%%%%%%%%%%%%%%%%%%%%%%%%%%%%%%%%%%%%%%%%%%%%%%%%%%%%%%%%%%%%%%%%%%%%%%%%%%%%%%%%%%%%%%
\section{Diagrammatic description of S-matrix elements for $\Gamma[A]$}
\label{sec2}
Based on the aforementioned discussion in Sec.~\ref{sec1}, here we would like to follow the work \cite{ghasemkhani-13} with a new approach, since we are interested in computation of the all-order noncommutative corrections to the Schwinger mass. However, as is well known, the standard computation of the NC phase factor for any diagram with an arbitrary number of loops is not an easy task. Thus, we are now presenting a stronger tool to produce this phase factor, which is technically easier than the use of the Feynman rule. Hence, the correct and simpler determination of the NC phase factor might be seen as an important feature of our S-matrix analysis. Moreover, in this approach, any complicated diagrams with an arbitrary number of loops are automatically generated.

We intend to study the ordinary S-matrix elements for the one-loop effective action $\Gamma[A]$. However, this action is completely different from the ordinary interacting theories such as QED, QCD, etc., since $\Gamma[A]$ includes a series with an infinite number of interaction terms; hence, caution is necessary.
As is well known, the S-matrix elements for a perturbative interacting theory described by $\mathcal{H}_{\rm int}$ are written in terms of the interaction part of the action as $S=\textsf{\large{T}}\exp{[i\int d^{4}x~\mathcal{H}_{\rm int}]}$ \cite{peskin}, where the symbol ``$\textsf{\large{T}}$'' denotes the time-ordering operator. Now, since we are interested in generating higher-order elements for the S-matrix for the case of interacting photons in the Schwinger model, we replace the usual interacting action $\mathcal{H}_{\rm int}$ by the new dynamics described by the effective action $\Gamma[A]$, so that the elements of the S-matrix are presented as
\begin{equation}
S=\textsf{\large{T}}e^{i\Gamma[A]}=1+\sum\limits_{n=1}^{\infty}S^{^{(n)}},
\label{eq:b1}
\end{equation}
where
\begin{equation}
S^{^{(n)}}=\frac{i^{n}}{n!}~\textsf{\large{T}}\Big(\Gamma[A]\Big)^{n}.
\label{eq:b2}
\end{equation}
Inserting Eq.~\eqref{eq:a8} into Eq.~\eqref{eq:b2} we obtain a more detailed expression as
\begin{align}
S&=1+\sum_{r}\int\prod_{j=1}^{r}d^{2}x_{j}~\textsf{\large{T}}\Big(A_{\mu_{1}}\left(x_{1}\right)\ldots A_{\mu_{r}}\left(x_{r}\right)\Big)~\Gamma^{\mu_{1}\ldots\mu_{r}}\left(x_{1},\ldots,x_{r}\right)\nonumber\\
&-\frac{1}{2}
\sum_{r,s}\int\prod_{j=1}^{r}d^{2}x_{j}\int\prod_{\ell=1}^{s}d^{2}y_{_{\ell}}~\textsf{\large{T}}
\Big(A_{\mu_{1}}\left(x_{1}\right)\ldots A_{\mu_{n}}\left(x_{r}\right)
A_{\nu_{1}}\left(y_{1}\right)\ldots A_{\nu_{s}}\left(y_{s}\right)\Big)\nonumber\\
&\times\Gamma^{\mu_{1}\ldots\mu_{r}}\left(x_{1},\ldots,x_{r}\right)
\Gamma^{\nu_{1}\ldots\nu_{s}}\left(y_{1},\ldots,y_{s}\right)
+\cdots.
\label{eq:b3}
\end{align}

We notice that the second term of \eqref{eq:b3} contains only one fermion loop while the next terms include more than one fermion loop. Therefore, in order to study \eqref{eq:b3} more precisely, we classify our analysis into two pieces: graphs with a single fermionic loop and those with multi-fermionic loop. These are respectively discussed below in Secs. \ref{sec2.1} and \ref{sec2.2}:

 %%%%%%%%%%%%%%%%%%%%%%%%%%%%%%%%%%%%%%%%%%%%%%%%%%%%%
\subsection{One-fermion-loop contribution to the photon self-energy sector}
\label{sec2.1}
 %%%%%%%%%%%%%%%%%%%%%%%%%%%%%%%%%%%%%%%%%%
This type of diagrams corresponds to $n=1$ of the series \eqref{eq:b1}, which is given by
\begin{equation}
S^{(1)}=i\sum\limits_{r=1}^{\infty}S^{(1,r)},
\label{eq:b4}
\end{equation}
where $S^{(1,r)}$ is defined by taking into account the expansion presented in Eq.~\eqref{eq:b3} as
\begin{equation}
S^{(1,r)}=\int\prod_{i=1}^{r}d^{2}x_{i}~\textsf{\large{T}}\Big(A_{\mu_{1}}(x_{1})\ldots A_{\mu_{r}}(x_{r})\Big)~\Gamma^{\mu_{1}\ldots \mu_{r}}(x_{1},\ldots,x_{r}).
\label{eq:b5}
\end{equation}

\begin{figure}[t]
\vspace{-1.3cm}
  \includegraphics[width=6.5cm,height=1.6cm]{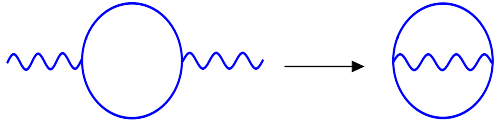}
     \centering  \caption{The Feynman graph description of $S^{(1,2)}$.}
     \label{fig2}
\end{figure}

Since our purpose is to compute quantum corrections to the photon self-energy, based on our previous discussion, we should consider an even value for $r$. So let us start with $r=2$
\begin{equation}
S^{(1,2)}=i\int d^{2}xd^{2}y~\textsf{T}\Big(A_{\mu}(x)A_{\nu}(y)\Big)~\Gamma^{\mu \nu}(x,y).
\label{eq:b-6}
\end{equation}
This expression can be simplified with the help of Wick's theorem \cite{peskin}. Writing down the Wick's expansion of $S^{(1,2)}$, we obtain
\begin{equation}
   S^{(1,2)}=i\int d^{2}xd^{2}y~:\left[A_{\mu}(x)A_{\nu}(y)
  +\underbracket[0.5pt]{A_{\mu}(x)A_{\nu}}(y)\right]:~\Gamma^{\mu \nu}(x,y),
  \label{eq:b7}
  \end{equation}
where we can write the tensor $\Gamma^{\mu \nu}(x,y)$ explicitly
\begin{equation}
\Gamma^{\mu \nu}(x,y)=-\frac{e^{2}}{2}\int\frac{d^{2}q}{(2\pi)^{2}}\frac{d^{2}\ell}
{(2\pi)^{2}}\frac{d^{2}p}{(2\pi)^{2}}(2\pi)^{2}
\delta(q+\ell)e^{i(q.x+\ell.y)}~\frac{tr\Big(\gamma^{\mu} (\displaystyle{\not}p+\displaystyle{\not}q) \gamma^{\nu}\displaystyle{\not}p\Big) }{(p+q)^{2} p^{2}}~e^{-\frac{i}{2}q\wedge
\ell},
\label{eq:b8}
\end{equation}
  here $:~:$ denotes the normal ordering operation. The noncommutative phase factor $e^{-\frac{i}{2}q \wedge \ell}$ is removed from \eqref{eq:b8} because of the energy-momentum conservation constraint $\delta\left(q+\ell\right)$. Consequently, the relevant Feynman graph in Fig.~\ref{fig2}, including a two-loop bubble diagram, is planar
\begin{equation}
   S^{(1,2)}_{\rm{bubble}}=i\int d^{2}xd^{2}y~{\cal{D}}_{\mu\nu}^{\tiny\mbox{(0)}}(x,y)
  ~\Gamma^{\mu \nu}(x,y).
 \label{eq:b9}
\end{equation}
Since the free propagator satisfies the relation ${\cal{D}}_{\mu\nu}^{\tiny\mbox{(0)}}(x,y)={\cal{D}}_{\mu\nu}^{\tiny\mbox{(0)}}(x-y)$, due to the
translational invariance \cite{pokorski}, we have
\begin{equation}
\int d^{2}xd^{2}y~{\cal{D}}_{\mu\nu}^{\tiny\mbox{(0)}}(x-y)e^{i(q.x+\ell.y)}
=\widetilde{\cal{D}}_{\mu\nu}^{\tiny\mbox{(0)}}(q,\ell)=
(2\pi)^{2}\delta(q+\ell)
\frac{-ig_{\mu\nu}}{q^{2}}.
\label{eq:b10}
\end{equation}
Substituting Eq.~\eqref{eq:b10} back into Eq.~\eqref{eq:b9}, and paying attention to the identity
\begin{equation}
\gamma^{\mu}\displaystyle{\not}p\gamma_{\mu}=(2-d)\displaystyle{\not}p,
\label{eq:b11}
\end{equation}
which vanishes in two dimensions, we arrive at the result that $ S^{ (1,2)}_{\rm{bubble}}=0$.
%%%%%%%%%%%%%%%%%%%%%%%%%%%%%%
%%%%%%%%%%%%%%%%%%%%%%%%%%%%%%

\begin{figure}[t]
\vspace{-1.5cm}\centering
  \includegraphics{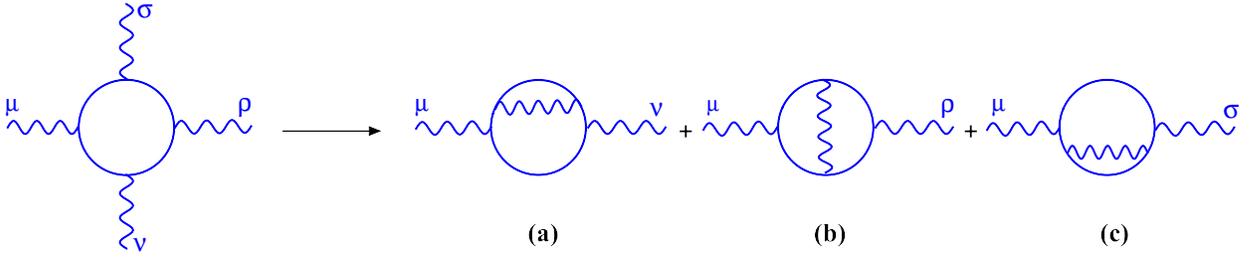}
    \caption{The Feynman graph description of $S^{(1,4)}$; the graphs (a),
(b), and (c) represent the two-loop contributions to the photon self-energy.}
    \label{four-line}
\end{figure}

Now, let us look at the next term $r=4$,
  \begin{equation}
  S^{(1,4)}=i\int \prod_{j=1}^{4} d^{2}x_{j}~\textsf{T}\Big(A_{\mu}(x_{1})A_{\nu}(x_{2})
  A_{\rho}(x_{3})A_{\sigma}(x_{4})\Big)~\Gamma^{\mu \nu\rho\sigma}(x_{1},\ldots,x_{4}).
  \label{eq:b12}
  \end{equation}
Applying Wick's theorem we find that
\begin{align}
     S^{(1,4)}&=i\int \prod_{j=1}^{4} d^{2}x_{j}:\Big[A_{\mu}(x_{1})A_{\nu}
  (x_{2})A_{\rho}(x_{3})A_{\sigma}(x_{4})+
  \underbracket[0.5pt]{A_{\mu}(x_{1})A_{\nu}}(x_{2})A_{\rho}(x_{3})A_{\sigma}(x_{4})
 \nonumber\\
 & +\underbracket[0.5pt]{A_{\mu}(x_{1})A_{\nu}(x_{2})A_{\rho}}(x_{3})A_{\sigma}(x_{4})
  +\underbracket[0.5pt]{A_{\mu}(x_{1})A_{\nu}(x_{2})A_{\rho}(x_{3})A_{\sigma}}(x_{4})+
  A_{\mu}(x_{1})\underbracket[0.5pt]{A_{\nu}(x_{2})A_{\rho}}(x_{3})A_{\sigma}(x_{4})
\nonumber\\
&+ A_{\mu}(x)\underbracket[0.5pt]{A_{\nu}(x_{2})A_{\rho}(x_{3})A_{\sigma}}(x_{4})+
 A_{\mu}(x_{1})A_{\nu}(x_{2})\underbracket[0.5pt]{A_{\rho}(x_{3})A_{\sigma}}(x_{4})
 \nonumber\\
 &+{\rm{other~possible~contractions}}\Big]:~\Gamma^{\mu \nu\rho\sigma}(x_{1},\ldots,x_{4}),
 \label{eq:b13}
  \end{align}
  where
  \begin{equation}
\Gamma^{\mu \nu\rho\sigma}(x_{1},\ldots,x_{4})=-\frac{e^{4}}{4}\int\prod_{j=1}^{4}\frac{d^{2}k_{j}}{(2\pi)^{2}}~(2\pi)^{2}~
\delta\big(\sum_{j=1}^{4}k_{j}\big)~e^{i\sum\limits_{j=1}^{4}k_{j}\cdot x_{j}}
e^{-\frac{i}{2}\sum\limits_{j<\ell}k_{j}\wedge
k_{\ell}}\Xi^{\mu \nu\rho\sigma},
\label{eq:b14}
\end{equation}
in which $\Xi^{\mu \nu\rho\sigma}$ is defined by inserting $n=4$ in Eq.~\eqref{eq:a10}.
Since we are interested in analyzing those higher-loop graphs contributing to the photon self energy, it is enough to consider terms with one contraction only. The relevant Feynman graphs are depicted in Fig.~\ref{four-line}. We observe that there are three independent graphs (a), (b), and (c), which are exactly the graphs appearing at order $e^{4}$ of QED S-matrix, contributing to the photon propagator at two-loop level.

Moreover, we intend to establish explicit expressions for the diagrams of the Fig.~\ref{four-line}, and show that their noncommutative phase factors are exactly the same as the ones obtained by direct use of the Feynman rules of the fermion-photon interaction in NC-QED. The corresponding S-matrix expressions for the diagrams of Fig.~\ref{four-line} are presented as
 \begin{align}
 S^{(1,4)}_{(a)} &  =i\int \prod_{j=1}^{4} d^{2}x_{j}:
 A_{\mu}(x_{1})A_{\nu}(x_{2}):{\cal{D}}_{\rho\sigma}^{(0)}(x_{3},x_{4})~\Gamma^{\mu \nu\rho\sigma}(x_{1},\ldots,x_{4}),
 \nonumber\\
  S^{(1,4)}_{(b)} &  =i\int \prod_{j=1}^{4} d^{2}x_{j}:
 A_{\mu}(x_{1})A_{\rho}(x_{3}):{\cal{D}}_{\nu\sigma}^{(0)}(x_{2},x_{4})~\Gamma^{\mu \nu\rho\sigma}(x_{1},\ldots,x_{4}),
 \nonumber\\
S^{(1,4)}_{(c)} &  =i\int \prod_{j=1}^{4} d^{2}x_{j}:
 A_{\mu}(x_{1})A_{\sigma}(x_{4}):{\cal{D}}_{\nu\rho}^{ (0)}(x_{2},x_{3})~\Gamma^{\mu \nu\rho\sigma}(x_{1},\ldots,x_{4}).
 \label{eq:b15}
  \end{align}
Besides, using the relations \eqref{eq:b10} and \eqref{eq:b14}, as well as the Fourier transform for the gauge fields, we obtain a simplified form for \eqref{eq:b15} as
\begin{align}
S^{(1,4)}_{(a)} & =\frac{e^{4}}{4}\int\frac{d^{2}q}{(2\pi)^{2}}\frac{d^{2}\ell}{(2\pi)^{2}}:
 \widetilde{A}_{\mu}(q)\widetilde{A}_{\nu}(-q):\frac{1}{\ell^{2}}~\widetilde{\Xi}^{\mu\nu}_{(a)},
 \nonumber\\
 S^{(1,4)}_{(b)}& =\frac{e^{4}}{4}\int\frac{d^{2}q}{(2\pi)^{2}}\frac{d^{2}\ell}{(2\pi)^{2}} :
 \widetilde{A}_{\mu}(q)\widetilde{A}_{\rho}(-q):
\frac{1 }{\ell^{2}}~\widetilde{\Xi}^{\mu\rho}_{ (b)}~e^{-iq\wedge \ell},
\nonumber\\
 S^{(1,4)}_{(c)} & =\frac{e^{4}}{4}\int\frac{d^{2}q}{(2\pi)^{2}}\frac{d^{2}\ell}{(2\pi)^{2}}:
 \widetilde{A}_{\mu}(q)\widetilde{A}_{\sigma}(-q):\frac{1}{\ell^{2}}~\widetilde{\Xi}^{\mu\sigma}_{ (c)},
 \label{eq:b16}
  \end{align}
where we have the following definition $\widetilde{\Xi}^{\mu\nu}_{(a)}=g_{\rho\sigma}\Xi^{\mu\nu\rho\sigma}_{(a)}$, $\widetilde{\Xi}^{\mu\rho}_{(b)}=g_{\nu\sigma}\Xi^{\mu\nu\rho\sigma}_{(b)}$, and $\widetilde{\Xi}^{\mu\sigma}_{(c)}=g_{\nu\rho}\Xi^{\mu\nu\rho\sigma}_{(c)}$, and such tensor expressions have the explicit form
  \begin{align}
 \widetilde{\Xi}^{\mu\nu}_{(a)} &= \int\frac{d^{2}p}{(2\pi)^{2}}~\frac{tr\Big(\gamma^{\mu}
(\displaystyle{\not}p+\displaystyle{\not}q)
\gamma^{\nu}(\displaystyle{\not}p+\displaystyle{\not}q+\displaystyle{\not}\ell)
\gamma^{\rho}(\displaystyle{\not}p+\displaystyle{\not}q)
\gamma_{\rho}\displaystyle{\not}p\Big)
}{(p+q)^{4}(p+q+\ell)^{2}p^{2}},
\nonumber\\
\widetilde{\Xi}^{\mu\rho}_{(b)} &=\int\frac{d^{2}p}{(2\pi)^{2}}~\frac{tr\Big(\gamma^{\mu}
(\displaystyle{\not}p+\displaystyle{\not}q)
\gamma^{\nu}(\displaystyle{\not}p+\displaystyle{\not}q+\displaystyle{\not}\ell)
\gamma^{\rho}(\displaystyle{\not}p+\displaystyle{\not}\ell)
\gamma_{\nu}\displaystyle{\not}p\Big)
}{(p+q)^{2}(p+q+\ell)^{2}(p+\ell)^{2}p^{2}},
\nonumber\\
\widetilde{\Xi}^{\mu\sigma}_{(c)} &= \int\frac{d^{2}p}{(2\pi)^{2}}~\frac{tr\Big(\gamma^{\mu}
(\displaystyle{\not}p+\displaystyle{\not}q)
\gamma^{\nu}\displaystyle{\not}p
\gamma_{\nu}(\displaystyle{\not}p+\displaystyle{\not}\ell)
\gamma^{\sigma}\displaystyle{\not}p\Big)
}{(p+q)^{2}(p+\ell)^{2}p^{4}},
 \label{eq:b17}
\end{align}
where $q$ is the momentum of the external gauge field, while $\ell$ and $p$ are momenta of the internal gauge and fermion fields, respectively.
 %%%%%%%%%%%%%%%%%%%%%%%%%%%%%%
%%%%%%%%%%%%%%%%%%%%%%%%%%%%%%

It is worth emphasizing that the presence of the energy-momentum conservation constraint, in the form of Dirac delta functions -- originating from Fourier transform of the free photon propagator in Eq.~\eqref{eq:b10} -- enables us to enforce these constraints upon the NC phase factor $e^{-\frac{i}{2}\sum\limits_{j<\ell}k_{j}\wedge k_{\ell}}$ and finally arrive at the correct phase factor. We notice, nonetheless, that the graphs (a) and (c) are planar and the graph (b) is nonplanar, as we
expected \cite{ghasemkhani-13}.

The integrals of Eq.~\eqref{eq:b17} can be analytically solved when written in the light-cone coordinates. To this end, we follow similar procedure to what was performed in \cite{ghasemkhani-13}, basically revising the integrals in terms of the light-cone coordinates. Then we should decompose them into partial fractions to reduce the degree of the denominator.
Moreover, we use the complex form of Green's theorem \cite{spiegel} and show that these integrals are vanishing; hence, we find that the two-loop corrections to the photon two-point function are absent.
This important result permits us to conclude that the Schwinger mass remains untouched at two-loop order. Further details on the application of the complex form of Green's theorem for our purpose are presented in appendix \ref{sec-appB}.

\begin{figure}[t]
\vspace{-1.0cm}\centering
  \includegraphics{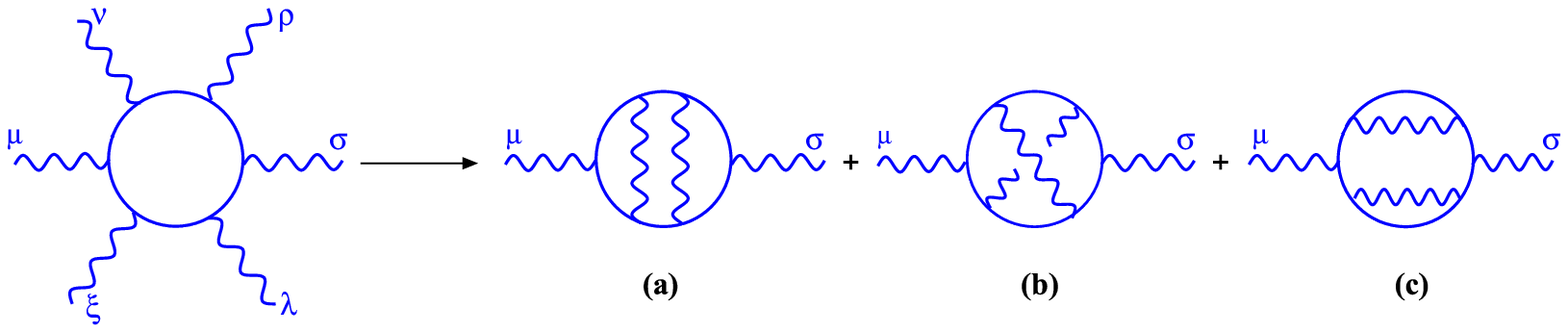}
   \caption{The Feynman graph description of $S^{(1,6)}$; the graphs (a), (b), and (c) represent the one-fermion-loop contributions to the photon self-energy at three-loop level.}
\label{six-line}
\end{figure}

In the last case of the graphs with a single fermionic loop, we take into account the contribution with $r=6$
  \begin{align}
   S^{(1,6)}&=i\int\prod_{j=1}^{6}d^{2}x_{j}~\textsf{T}\Big(A_{\mu}(x_{1})
   A_{\nu}(x_{2})A_{\rho}(x_{3})A_{\sigma}(x_{4})
  A_{\lambda}(x_{5})A_{\xi}(x_{6})\Big)~\Gamma^{\mu\nu\rho\sigma\lambda\xi}(x_{1},\ldots,x_{6})
    \nonumber\\
    &=i\int\prod_{j=1}^{6}d^{2}x_{j}~~\Gamma^{\mu \nu\rho\sigma\lambda\xi}(x_{1},\ldots,x_{6}):\Big[A_{\mu}(x_{1})A_{\nu}(x_{2})A_{\rho}(x_{3})A_{\sigma}(x_{4}) A_{\lambda}(x_{5})A_{\xi}(x_{6})
  \nonumber\\
  &+   A_{\mu}(x_{1})\overbracket[0.5pt]{A_{\nu}(x_{2})\underbracket[0.5pt]{A_{\rho}(x_{3})A_{\sigma}(x_{4}) A_{\lambda}}(x_{5})A_{\xi}}(x_{6})
  + A_{\mu}(x_{1})\overbracket[0.5pt]{A_{\nu}(x_{2})\underbracket[0.5pt]{A_{\rho}(x_{3})A_{\sigma}(x_{4}) A_{\xi}}(x_{6})A_{\lambda}}(x_{5})
   \nonumber\\
   &+A_{\mu}(x_{1})\overbracket[0.5pt]{A_{\nu}(x_{2})A_{\rho}}(x_{3})A_{\sigma}(x_{4})    \underbracket[0.5pt]{A_{\lambda}(x_{5})A_{\xi}}(x_{6})+\rm{other~possible~contractions}\Big]:.
   \label{eq:b18}
  \end{align}
  Again, we have focused on the terms with two external legs, after performing Wick's expansion. The contributing graphs are depicted in Fig.~\ref{six-line}. Now, we want to obtain the related forms of these three contracted terms,
  \begin{align}
 S^{(1,6)}_{(a)} &=
 i\int\prod_{j=1}^{6}d^{2}x_{j}:
    A_{\mu}(x_{1})A_{\sigma}(x_{4})
 :{\cal D}_{\nu\xi}^{(0)}(x_{2},x_{6})~{\cal D}_{\rho\lambda}^{(0)}(x_{3},x_{5})~\Gamma^{\mu \nu\rho\sigma\lambda\xi}(x_{1},\ldots,x_{6}),
 \nonumber\\
  S^{(1,6)}_{(b)} &=
 i\int\prod_{j=1}^{6}d^{2}x_{j}:
    A_{\mu}(x_{1})A_{\sigma}(x_{4})
 :{\cal D}_{\nu\lambda}^{(0)}(x_{2},x_{5})~{\cal D}_{\rho\xi}^{(0)}(x_{3},x_{6})~\Gamma^{\mu \nu\rho\sigma\lambda\xi}(x_{1},\ldots,x_{6}),
 \nonumber\\
 S^{(1,6)}_{(c)}&=
 i\int\prod_{j=1}^{6}d^{2}x_{j}:
    A_{\mu}(x_{1})A_{\sigma}(x_{4})
 :{\cal D}_{\nu\rho}^{(0)}(x_{2},x_{3})~{\cal D}_{\xi\lambda}^{(0)}(x_{5},x_{6})~\Gamma^{\mu \nu\rho\sigma\lambda\xi}(x_{1},\ldots,x_{6}),
    \label{eq:b19}
  \end{align}
   where
 \begin{equation}
 \Gamma^{\mu\nu\rho\sigma\lambda\xi}(x_{1},\ldots,x_{6})=-\frac{e^{6}}{6}
\int\prod_{j=1}^{6}\frac{d^{2}k_{j}}{(2\pi)^{2}}~(2\pi)^{2}
\delta(\sum_{j=1}^{6}k_{j})~e^{i\sum\limits_{j=1}^{6}k_{j}\cdot x_{j}}
e^{-\frac{i}{2}\sum\limits_{j<\ell}k_{j}\wedge
k_{\ell}}\Xi^{\mu \nu\rho\sigma\lambda\xi},
\label{eq:b20}
 \end{equation}
and that $\Xi^{\mu \nu\rho\sigma\lambda\xi}$ is determined by inserting $n=6$ into Eq.~\eqref{eq:a10}. After carrying out some computational steps, we find out that graphs (a) and (b) are nonplanar and the graph (c) is planar. The relevant results are listed as follows
\begin{align}
 S^{(1,6)}_{(a)} & =\frac{e^{6}}{6}\int\frac{d^{2}q}{(2\pi)^{2}}\frac{d^{2}\ell}{(2\pi)^{2}}\frac{d^{2}k}{(2\pi)^{2}} :
 \widetilde{A}_{\mu}(q)\widetilde{A}_{\sigma}(-q):
\frac{1}{\ell^{2}k^{2}}~\widetilde\Xi^{\mu \sigma}_{_{(a)}}~e^{-iq\wedge \left(\ell+k\right)} \nonumber\\
S^{(1,6)}_{(b)} & =\frac{e^{6}}{6}\int\frac{d^{2}q}{(2\pi)^{2}}\frac{d^{2}\ell}{(2\pi)^{2}}\frac{d^{2}k}{(2\pi)^{2}} :
 \widetilde{A}_{\mu}(q)\widetilde{A}_{\sigma}(-q):
\frac{1}{\ell^{2}k^{2}}~\widetilde\Xi^{\mu \sigma}_{_{(b)}}~e^{-i\left[q\wedge \left(\ell+k\right)+\ell\wedge k\right]}
,\nonumber\\
S^{(1,6)}_{(c)} & =\frac{e^{6}}{6}\int\frac{d^{2}q}{(2\pi)^{2}}\frac{d^{2}\ell}{(2\pi)^{2}}\frac{d^{2}k}{(2\pi)^{2}} :
 \widetilde{A}_{\mu}(q)\widetilde{A}_{\sigma}(-q):
\frac{1}{\ell^{2}k^{2}}~\widetilde\Xi^{\mu \sigma}_{_{(c)}},
\label{eq:b21}
  \end{align}
 where $(\ell,k)$ are the momenta of the internal gauge fields and we have also defined
 \begin{equation}
  \widetilde\Xi^{\mu \sigma}_{(a)}=g_{\nu\xi}g_{\rho\lambda}\Xi^{\mu \nu\rho\sigma\lambda\xi}_{(a)},\quad
  \widetilde\Xi^{\mu \sigma}_{(b)}=g_{\nu\lambda}g_{\rho\xi}\Xi^{\mu \nu\rho\sigma\lambda\xi}_{(b)},\quad
     \widetilde\Xi^{\mu \sigma}_{(c)}=g_{\nu\rho}g_{\xi\lambda}\Xi^{\mu \nu\rho\sigma\lambda\xi}_{(c)},
     \label{eq:b22}
 \end{equation}
which are explicitly described as
%%%%%%%%%%%%%%%%%%%%%%%%%%%%%%
%%%%%%%%%%%%%%%%%%%%%%%%%%%%%%
\begin{align}
\widetilde\Xi^{\mu \sigma}_{(a)} &=\int\frac{d^{2}p}{(2\pi)^{2}}
~\frac{{\cal N}^{\mu \sigma}_{(a)}}
{p^{2}(q+p)^{2}(q+p+\ell)^{2}(q+p+\ell+k)^{2}(p+\ell+k)^{2}(p+\ell)^{2}},
\nonumber\\
%%%%%
\widetilde\Xi^{\mu \sigma}_{(b)} &=\int\frac{d^{2}p}{(2\pi)^{2}}
~\frac{{\cal N}^{\mu\sigma}_{(b)}}
{p^{2}(q+p)^{2}(q+p+\ell)^{2}(q+p+\ell+k)^{2}(p+\ell+k)^{2}(p+k)^{2}},
\nonumber\\
%%%%%%%%%%%%%%
\widetilde\Xi^{\mu \sigma}_{(c)} &=\int\frac{d^{2}p}{(2\pi)^{2}}
~\frac{{\cal N}^{\mu \sigma}_{(c)}}
{p^{4}(q+p)^{2}(q+p+\ell)^{2}(p+q)^{2}(p+k)^{2}},
\label{eq:b22-new}
\end{align}
with
\begin{align}
{\cal N}^{\mu \sigma}_{(a)} &=tr\Big(\gamma^{\mu}
(\displaystyle{\not}q+\displaystyle{\not}p)\gamma^{\rho}(\displaystyle{\not}q+\displaystyle{\not}p+
\displaystyle{\not}\ell)\gamma^{\nu}(\displaystyle{\not}q+\displaystyle{\not}p+\displaystyle{\not}\ell+\displaystyle{\not}k)
\gamma^{\sigma}(\displaystyle{\not}p+\displaystyle{\not}\ell+\displaystyle{\not}k)\gamma_{\nu}(\displaystyle{\not}p+
\displaystyle{\not}\ell)\gamma_{\rho}
\displaystyle{\not}p\Big),
\nonumber\\
{\cal N}^{\mu \sigma}_{(b)} &=tr\Big(\gamma^{\mu}
(\displaystyle{\not}p+\displaystyle{\not}q)\gamma^{\rho}
(\displaystyle{\not}p+\displaystyle{\not}q+\displaystyle{\not}\ell)\gamma^{\nu}
(\displaystyle{\not}p+\displaystyle{\not}q+\displaystyle{\not}\ell+\displaystyle{\not}k)
\gamma^{\sigma}(\displaystyle{\not}p+\displaystyle{\not}\ell+\displaystyle{\not}k )
\gamma_{\rho}(\displaystyle{\not}p+\displaystyle{\not}k)\gamma_{\nu}\displaystyle{\not}p\Big),
 \nonumber\\
{\cal N}^{\mu \sigma}_{(c)} &=tr\Big(\gamma^{\mu}
(\displaystyle{\not}p+\displaystyle{\not}q)
\gamma_{\rho}(\displaystyle{\not}q+\displaystyle{\not}p+\displaystyle{\not}\ell)
\gamma^{\rho}(\displaystyle{\not}p+\displaystyle{\not}q)\gamma^{\sigma}
\displaystyle{\not}p\gamma^{\nu}(\displaystyle{\not}p+\displaystyle{\not}k)\gamma_{\nu}\displaystyle{\not}p\Big).
\label{eq:b23-new}
\end{align}

It is worth noting that the NC phase factors of the set of expressions \eqref{eq:b16} and \eqref{eq:b21} are exactly the same as those found by making direct use of the Feynman rule of the NC-QED Lagrangian in Ref.~\cite{ghasemkhani-13}.

 Regarding the analytical computation of the integrals over the fermion-loop momentum $p$ in \eqref{eq:b22-new}, we follow exactly the approach previously discussed in the case of $S^{(1,4)}$. Proceeding in this way, we find out that $\widetilde\Xi^{\mu \sigma}_{(a)}$, $\widetilde\Xi^{\mu \sigma}_{(b)}$, and $\widetilde\Xi^{\mu \sigma}_{(c)}$ are all vanishing (see further details in appendix~\ref{sec-appB}). Therefore, this means that there are no corrections to the photon two-point function and to the Schwinger mass at three-loop level.

To conclude our discussion concerning diagrams of a single fermionic loop, we consider the generic term $ S^{(1,r)}$, which is proportional to $e^{r}$, and that after Wick's expansion includes a $r-2$ number of photon propagators and produces different one-loop graphs contributing to photon self-energy at order $r$. The appearance of a $r-2$ number of photon propagators in such a contribution, indeed generates a $r-2$ number of Dirac delta functions that render to easily determine the NC phase factor.

However, it is important to emphasize that the NC phase factors associated to all the nonplanar graphs are independent from the fermion-loop momentum, depending only on the external and internal photons momenta.
Hence, for the sake of simplicity, we first carry out the integral over the fermion-loop momentum in the light-cone coordinates. Performing some algebraic manipulations and using the complex form of Green's theorem eventually permit us to realize that the amplitudes of these graphs are zero (in the same sense as we have discussed for the previous cases).

It should be emphasized that all of the diagrams discussed here include one-fermionic-loop only and in what follows the multi-fermionic-loop contributions are considered as well, and we want to highlight that this approach can easily give us the exact value of the NC phase factor with increasing the number of fermion loops in a simpler fashion than the use of Feynman rules.

%%%%%%%%%%%%%%%%%%%%%%%%%%%%%%%%%%%%%%%%%%%%%%%%%%%%%%%%%%%%%%%%%%%%%%%%%%%%%%%%%%%%%%%%%%%%%%%%%%%%%%%%%%%%%%%%%%%%%%%%%%%%%
\subsection{Multi-fermion-loop contribution to the photon self-energy sector}
\label{sec2.2}
%%%%%%%%%%%%%%%%%%%%%%%%%%%%%%%%%%%%%%%%%%%%%%%%%%%%%%%%%%%%%%%%%%%%%%%%%%%%%%%%%%%%%%%%%%%%%%%%%%%%%%%%%%%%%%%%%%%%%%%%%%%%%%
In this subsection we intend to establish the contribution of graphs with more than one fermionic loop, corresponding to $n\geq 2$ in Eq.~\eqref{eq:b1}, to the photon self-energy part. First, let us start with $n=2$, including two-fermion-loop diagrams
\begin{equation}
S^{(2)}=-\frac{1}{2}\sum\limits_{r=1}^{\infty}\sum\limits_{s=1}^{\infty}S^{(2,r,s)},
\label{eq:b23}
\end{equation}
with $r$ and $s$ running over the first and second fermionic loop respectively, as can be seen from
\begin{align}
S^{(2,r,s)}&=\int\prod_{i=1}^{r}d^{2}x_{i}\int\prod_{j=1}^{s}d^{2}y_{j}~\textsf{T}\Big(A_{\mu_{1}}(x_{1})\ldots A_{\mu_{r}}(x_{r})
A_{\nu_{1}}(y_{1})\ldots A_{\nu_{s}}(y_{s})
\Big)\nonumber\\
&\times \Gamma^{\mu_{1}\ldots \mu_{r}}(x_{1},\ldots,x_{r})~\Gamma^{\nu_{1}\ldots \nu_{s}}(y_{1},\ldots,y_{s}).
\label{eq:b24}
\end{align}
%%%%%%%%%%%%%%%%%%%%%%%%%%%%%%
\begin{figure}[t]
\vspace{-1.5cm}\centering
  \includegraphics{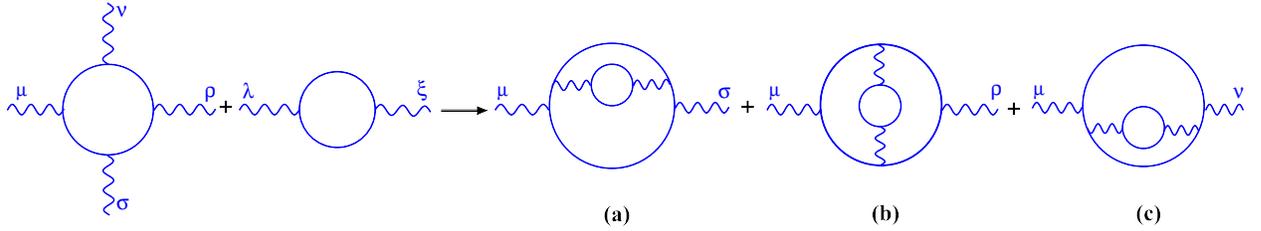}
  \caption{The Feynman graph description of $S^{(2,4,2)}$; the graphs (a),(b), and (c) represent the two-fermion-loop contributions to the photon self-energy at three-loop level.}
\label{Mixed-graph}
\end{figure}
%%%%%%%%%%%%%%%%%%%%%%%%%%%%%%
In the case in which $n\geq 2$, when the Wick's theorem is used to expand the $\textsf{T}$ product it produces a number of connected and disconnected diagrams. This is in contrast to the $n=1$ case, where it included solely connected graphs. However, since we are interested in discussing the noncommutative corrections to the massive pole of the gauge field propagator, we retain only the connected graphs in our analysis. Now, for illustration, we choose $r=4$ and $s=2$ in Eq.~\eqref{eq:b24}
\begin{align}
  S^{(2,4,2)}&=-\frac{1}{2}\int\prod_{i=1}^{4}d^{2}x_{i}\int\prod_{j=1}^{2}d^{2}y_{j}~\textsf{T}\Big(A_{\mu}(x_{1})A_{\nu}(x_{2})A_{\rho}(x_{3})A_{\sigma}(x_{4})
  A_{\lambda}(y_{1})A_{\xi}(y_{2})\Big)\nonumber\\
  &\times\Gamma^{\mu\nu\rho\sigma}(x_{1},\ldots,x_{4})
  ~\Gamma^{\lambda\xi}(y_{1},y_{2}).
  \label{eq:b25}
  \end{align}
  Making use of Wick's theorem on Eq.~\eqref{eq:b25}, we find
  \begin{align}
    S^{(2,4,2)}&=-\frac{1}{2}\int\prod_{i=1}^{4}d^{2}x_{i}\int\prod_{j=1}^{2}d^{2}y_{j}~\Gamma^{\mu\nu\rho\sigma}(x_{1},\ldots,x_{4})~
  \Gamma^{\lambda\xi}(y_{1},y_{2})\nonumber\\
  &\times:     \Big[A_{\mu}(x_{1})A_{\nu}(x_{2})A_{\rho}(x_{3})A_{\sigma}(x_{4}) A_{\lambda}(y_{1})A_{\xi}(y_{2})   +
  A_{\mu}(x_{1})\overbracket[0.5pt]{A_{\nu}(x_{2})A_{\rho}(x_{3})\underbracket[0.5pt]{A_{\sigma}(x_{4}) A_{\lambda}}(y_{1})A_{\xi}}(y_{2})
  \nonumber\\
  &+
   A_{\mu}(x_{1})\overbracket[0.5pt]{A_{\nu}(x_{2})\underbracket[0.5pt]{A_{\rho}(x_{3})A_{\sigma}(x_{4}) A_{\lambda}}(y_{1})A_{\xi}}(y_{2})
  %%%%%%%
  +
   A_{\mu}(x_{1})A_{\nu}(x_{2})\overbracket[0.5pt]{A_{\rho}(x_{3})\underbracket[0.5pt]{A_{\sigma}(x_{4}) A_{\lambda}}(y_{1})A_{\xi}}(y_{2})
  \nonumber\\
  &+ \rm{other~possible~contractions}\Big]:.
  \label{eq:b26}
  \end{align}
Now, let us consider the first term of Eq.~\eqref{eq:b26}, corresponding to graph (b) of Fig.~\ref{Mixed-graph}
\begin{align}
  S^{(2,4,2)}_{(a)}&=-\frac{1}{2}\int\prod_{i=1}^{4}d^{2}x_{i}\int\prod_{j=1}^{2}d^{2}y_{j}:
  A_{\mu}(x_{1})A_{\sigma}(x_{4}): \Gamma^{\mu\nu\rho\sigma}{\cal D}_{\nu\rho}^{(1)},
 \nonumber\\
 S^{(2,4,2)}_{(b)}&=-\frac{1}{2}\int\prod_{i=1}^{4}d^{2}x_{i}\int\prod_{j=1}^{2}d^{2}y_{j}:
  A_{\mu}(x_{1})A_{\rho}(x_{3}):\Gamma^{\mu\nu\rho\sigma}{\cal D}_{\nu\sigma}^{(1)},
  \nonumber\\
    S^{(2,4,2)}_{(c)}&=-\frac{1}{2}\int\prod_{i=1}^{4}d^{2}x_{i}\int\prod_{j=1}^{2}d^{2}y_{j}:
  A_{\mu}(x_{1})A_{\nu}(x_{2}):\Gamma^{\mu\nu\rho\sigma}{\cal D}_{\sigma\rho}^{(1)},
  \label{eq:b27}
  \end{align}
    where we have defined
  \begin{align}
  {\cal D}_{\nu\rho}^{(1)} &={\cal D}_{\nu\lambda}^{(0)}(x_{2}-y_{1})\Gamma^{\lambda\xi}(y_{1},y_{2}){\cal D}_{\xi\rho}^{(0)}(y_{2}-x_{3}),\nonumber\\
  {\cal D}_{\nu\sigma}^{(1)} &={\cal D}_{\nu\lambda}^{(0)}(x_{2}-y_{1})\Gamma^{\lambda\xi}(y_{1},y_{2}){\cal D}_{\xi\sigma}^{(0)}(y_{2}-x_{4}),\nonumber\\
  {\cal D}_{\sigma\rho}^{(1)} &={\cal D}_{\sigma\lambda}^{(0)}(x_{4}-y_{1})\Gamma^{\lambda\xi}(y_{1},y_{2}){\cal D}_{\xi\rho}^{(0)}(y_{2}-x_{3}).
  \label{eq:b28}
 \end{align}

Here we have introduced ${\cal D}_{\nu\rho}^{(1)}$ by simplicity, where it corresponds to the one-loop corrected photon propagator due to a fermion loop, the gauge and ghost loops disappear in two-dimensional light-cone coordinates. The detailed forms of $\Gamma^{\mu\nu\rho\sigma}$ and $ \Gamma^{\lambda\xi}$ were illustrated in Eqs.~\eqref{eq:b14} and \eqref{eq:b8} including two energy-momentum conservation constraints $\delta(\sum\limits_{i=1}^{4}k_{i})$ and $\delta(\sum\limits_{j=1}^{2}\ell_{j})$, respectively.
Notice that these constraints along with the new ones: $\delta\left(k_{2}+\ell_{1}\right)\delta\left(k_{3}+\ell_{2}\right)$, $\delta\left(k_{2}+\ell_{1}\right)\delta\left(k_{4}+\ell_{2}\right)$, and
$\delta\left(k_{4}+\ell_{1}\right)\delta\left(k_{3}+\ell_{2}\right)$, obtained from Eq.~\eqref{eq:b28}, allow us to specify the value of the NC phase factors associated with graphs (a), (b), and (c), respectively.
Based on these considerations, we can conclude that graphs (a) and (c) are planar while graph (b) is nonplanar with the NC phase $e^{i\ell_{1}\wedge k_{1}}$.
It should be emphasized that the diagrams of Fig.~\ref{Mixed-graph}, which are proportional to $e^{6}$, give a three-loop contribution to the photon two-point function and so they should be added to the three-loop contributions coming from the one-fermionic loop, i.e., Eqs.~\eqref{eq:b21}.
However, we will show next that, similar to \eqref{eq:b21}, the contribution \eqref{eq:b27} vanishes and we do not find again any corrections at this order too.

\begin{figure}[t]
\vspace{-1.5cm}\centering
  \includegraphics[width=8.1cm,height=2.9cm]{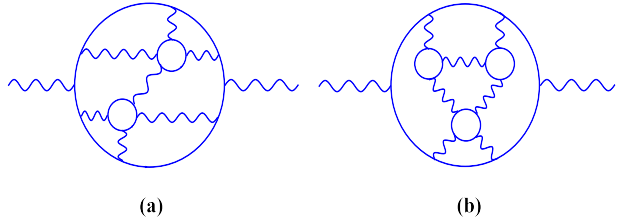}
  \caption{Multi-fermion-loop diagrams: (a) belongs to $S^{(3,8,4,4)}$ and (b) belongs to $S^{(4,6,3,3,4)}$.}
\label{Multi-loop}
\end{figure}

To conclude the study of the diagrams of the multi-fermion-loop type, we present a general relation for a graph with an arbitrary number of fermion loops $(n\geq 2)$, using Eqs.~\eqref{eq:b1}-\eqref{eq:b3}
\begin{equation}
 S^{(n)}=\frac{i^{n}}{n!}\sum_{r_{1},\ldots,r_{n}}S^{(n,r_{1},\ldots,r_{n})},
\end{equation}
where $S^{(n,r_{1},\ldots,r_{n})}$ refers to a set of graphs with $n$ fermion loops at order $e^{\sum\limits_{i=1}^{n}r_{i}}$, in which $r_{1},\ldots,r_{n}$ indicate the number of the external gauge fields associated with each fermion loop.
For example, two multi-fermion-loop graphs have been depicted in Fig.~\ref{Multi-loop}. As it can be easily checked, both graphs (a) and (b) are at order $e^{16}$ and belong to $S^{(3,8,4,4)}$ and $S^{(4,6,3,3,4)}$, respectively.
Also, the first graph includes 3 fermion loops with 8, 4, and 4 external legs,
while the second one includes 4 fermion loops with 6, 3, 3, and 4 external legs. However, our analysis shows that all of these higher-loop corrections are vanishing in two dimensions.
%%%%%%%%%%%%%%%%%%%%%%%%%%%%%%
%%%%%%%%%%%%%%%%%%%%%%%%%%%%%%%%%%%%%%%%%%%%%%%%%%%%%%%%%%%%%%%%%%%%%%%%%%%%%%%%%%%%%%%%%%%%%
\section{Generalization to all translation-invariant noncommutative star products}
\label{sec4}
%%%%%%%%%%%%%%%%%%%%%%%%%%%%%%%%%%%%%%%%%%%%%%%%%%%%%%%%%%%%%%%%%%%%%%%%%%%%%%%%%%%%%%%%%%%%%
It seems worth stressing at this point that simplifying the properties of the Moyal product was extensively used to extract all of above achievements.
However, one may as well be interested in analyzing the present field theory model when the usual Moyal product is replaced by another noncommutative star product. So it is important to realize that the obtained results and statements are valid independently of the chosen noncommutative star product.

To answer the previous statement, we shall restrict ourselves to translation-invariant star
products. In fact, due to the Noether's theorem, in order to preserve the energy-momentum conservation law, it is mandatory to restrict noncommutative star products to those which do not depend explicitly on the coordinates of the chosen spacetime frame, which, in return are commonly referred to as translation-invariant noncommutative star products\cite{Lizzi,Galluccio}.

Essentially, using the theory of $\alpha^*$-cohomology \cite{Lizzi,Galluccio} to characterize the quantum behaviors of translation-invariant noncommutative quantum field theories, has shown that \cite{Varshovi-1, Varshovi-2}

\begin{enumerate}[(a)]

\item The whole quantum behavior of two translation-invariant noncommutative versions of a specific renormalizable quantum field theory with two $\alpha^*$-cohomologous star products, say $\star_1$ and $\star_2$, coincide precisely. In fact, the equality of scattering matrices of theories ${\cal{L}}_{\star_1}$ and ${\cal{L}}_{\star_2}$ holds for any renormalizable theory $\mathcal{L}$ and for each $\star_1 \sim \star_2$ \cite{Varshovi-1}.

\item Any given translation-invariant noncommutative is $\alpha^*$-cohomologous to a unique Moyal star product \cite{Varshovi-2}.

\end{enumerate}

Therefore, it follows that from assertion (a) the calculated scattering matrix of noncommutative Moyal QED$_2$ holds exactly the same for any translation-invariant noncommutative QED$_2$, but only to the class of star products that are $\alpha^*$-cohomologous to that of Moyal.
Moreover, since there is a unique Moyal product in two-dimensional spacetime, thus according to (b), the given formula \eqref{eq:b3} definitely provides the scattering matrix for any given noncommutative energy-momentum preserving version of QED$_2$.
In other words, according to \cite{Varshovi-2}, for a two-dimensional spacetime, which presents only one $\alpha^*$-cohomology class of noncommutative translation-invariant star products, the calculated scattering matrix \eqref{eq:b3} is the only energy-momentum preserving modification of that obtained in standard Euclidean QED$_2$.

Actually this result seems to be rather intriguing, since the following two critical points hold only for the case of Moyal noncommutative structures:

\begin{enumerate}

\item It is well known that the Moyal product between two functions disappears under the integration. In fact, in the definition \eqref{eq:a8}, this important property of the Moyal star product has been used extensively throughout our analysis, where the bosonic fields have been put next to each other before $\Gamma^{\mu_1\ldots \mu_m}(x_1,\ldots,x_m)$ with no appearance of the noncommutative star product. Nonetheless, although Eq.~\eqref{eq:a6} holds for any general noncommutative star product, generally labeled by $\star$, in order to encompass such a general discussion, \eqref{eq:a8} should be modified as
\begin{equation} \label{modified (2.8)}
\Gamma_{\star}[A]=\sum_{n=1}^{\infty}\int \prod_{i=1}^nd^2x_i~\left( A_{\mu_1}(x_1)~A_{\mu_2}(x_2)~\ldots~A_{\mu_n}(x_n)\right)\star_1 \star_2\ldots\star_n~\Gamma^{\mu_1...\mu_n}(x_1,\ldots,x_n),
\end{equation}
where $\star_i$ is the star product for variable $x_i$, $i=1,\ldots,n$.

\item The phase factor of the noncommutative product in formula \eqref{eq:a-9} is essentially independent of the momentum of fermions due to energy-momentum conservation constraints present in the Feynman graphs. As we have seen above, this significant property helped us to integrate over the fermionic momentum with no appearance of phase factor; this is rather similar to the case of standard commutative QED, in which the two-dimensional Euclidean spacetime leads to cancellation of the whole momentum integral (see appendix \ref{sec-appB} for further detail).

\end{enumerate}

Without the validity of these two points it seems somehow impossible to prove the final formula of scattering matrix in the case of Moyal product holds for a general translation-invariant noncommutative star product.
However, it must be emphasized that this interesting achievement is due to the cohomological point of view developed in Refs.~\cite{Varshovi-1,Varshovi-2} to characterize translation-invariant noncommutative field theories.
To explain this in a simple reasoning, it is enough to assert that any translation-invariant noncommutative star product can be converted to a unique Moyal product through of a redefinition of fields, due to the ``Hodge theorem'' in $\alpha^*$-cohomology \cite{Varshovi-1}.

Particularly, the mentioned redefinition is basically independent of the chosen type of fields and spacetime dimensionality.
Therefore, this redefinition can be absorbed thoroughly by the path-integral measure and thus it never actually contributes to the scattering matrix of the theory.
This is evidently a direct consequence of the ``quantum equivalence theorem'' which states that the whole physical effects of a (renormalizable) translation-invariant noncommutative quantum field theory are preserved through a $\alpha^*$-cohomology class \cite{Varshovi-1}.

Let us establish the aforementioned results with some more details by means of a noncohomological approach. Here we need some conventions. In what follows, the Moyal product is represented with $\star$, while $\star'$ is used for general translation-invariant star product. Moreover, classical gauge field is shown with simple notation of $A$ whereas its second quantized version is denoted by $\hat A$.

Because of Eq.~\eqref{eq:b1} the scattering matrix for star product $\star'$ is
\begin{equation} \label{modified (3.1)}
S_{\star'}=\textsf{T}e^{i\Gamma_{\star'}[\hat A]}=1+\sum_{n=1}^{\infty}\frac{{i^n}}{{n!}}\textsf{T}\left(\Gamma_{\star'}[\hat A]\right)^n,
\end{equation}
where $\Gamma_{\star'}[A]$ is defined via \eqref{modified (2.8)} for star product $\star'$.
Actually, if the effective action $\Gamma_{\star'}[A]$ is given for a classical field $A$, the scattering matrix $S_{\star'}$ is worked out by replacing $A$ by its second quantized form $\hat A$, which is then expanded in terms of creation and annihilation operators over the Fock space. Therefore, if we show that
\begin{equation} \label {Gamma A to Gamma A'}
\Gamma_{\star'}[A]=\Gamma_{\star}[A'].
\end{equation}
holds for some classical field $A'$ which is a transformed version of $A$ as
\begin{equation}
A'_{\mu}(x)=\int \frac{{d^2p}}{{(2\pi )^2}}~\tilde{A}_{\mu}(p)~e^{\beta(p)}~e^{-ip.x},
 \label{transformation}
\end{equation}
where $\tilde{A}_{\mu}(p)$ is the Fourier transform of $A_{\mu}(x)$ and $\beta:\mathbb{R}^2 \to \mathbb{R}$ is a smooth function. Then, we readily find that $\Gamma_{\star'}[\hat{A}]=\Gamma_{\star}[\hat{A}]$, which consequently leads to $S_{\star'}=S_{\star}$ due to the series equation \eqref{modified (3.1)}, as we claimed above. \footnote{In fact, this is a direct consequence of the normalization of fields via the LSZ reduction formula.}

It is worth recalling that a more precise expression of Eq.~\eqref{eq:a5} for $\Gamma_{\star'}[A]$ is defined as
\begin{equation} \label{Gamma_star}
e^{i\Gamma_{\star'}[A]}=\frac{{\int {\cal{D}}\overline \psi {\cal{D}}\psi ~e^{i{\cal{S}}_{f,\star'}(\overline \psi,\psi,A)}}}{{\int {\cal{D}}\overline \psi {\cal{D}}\psi ~e^{i{\cal{S}}_{f,\star'}(\overline \psi,\psi,0)}}},
\end{equation}
for a classical gauge field $A$, and we have defined
\begin{equation} \label {Action_star}
{\cal{S}}_{f,\star'}(\overline \psi,\psi,A)=\int d^2x ~\left[\overline \psi \star' i\gamma^{\mu}\partial_{\mu}\psi-e\overline \psi \star' A_{\mu}\star' \psi-m\overline \psi \star' \psi \right].
\end{equation}

In Refs.~\cite{Varshovi-1,Varshovi-2}, it has been shown that for any $\star'$ there exists an appropriate function $\beta:\mathbb{R}^2 \to \mathbb{R}$ which for any set of functions, say $\{f_1,\ldots,f_n\}$, implies
\begin{equation} \label {star' to star}
\int d^2x~f_1\star'f_2\star'\ldots\star'f_n=\int d^2x~f'_1\star f'_2\star\ldots\star f'_n,
\end{equation}
where the transformed functions $\{f'_1,\ldots,f'_n\}$ are given as
\begin{equation}
f'(x)=\int \frac{{d^2p}}{{(2\pi )^2}}~\tilde{f}(p)~e^{\beta(p)}~e^{-ip.x},
 \label{transformation}
\end{equation}
where $\tilde{f}$ is the Fourier transform of $f$. However, applying the identities ,Eqs.~\eqref{star' to star} and \eqref{transformation}, into \eqref{Action_star} we conclude that
\begin{equation}
{\cal{S}}_{f,\star'}(\overline \psi,\psi,A)={\cal{S}}_{f,\star}(\overline \psi',\psi',A').
 \label{Action_star-2}
\end{equation}

It is not difficult to see that ${\cal{D}}\overline \psi {\cal{D}}\psi=K{\cal{D}}\overline \psi' {\cal{D}}\psi'$ for some constant $K$. \footnote{In fact, this can be easily seen for a Fourier transformed path-integral measure. First notice that ${\cal{D}} \tilde{\overline \psi} {\cal{D}} \tilde{\psi}=\prod\limits_{p\in \mathbb{R}^2} d \tilde{\overline \psi} (p)d \tilde{\psi} (p)=\prod\limits_{p\in \mathbb{R}^2} e^{-2\beta(p)}~d \tilde{\overline \psi'} (p)d \tilde{\psi'} (p)=\prod\limits_{q\in \mathbb{R}^2} e^{-2\beta(q)}~\times~\prod\limits_{p \in \mathbb{R}^2}d \tilde{\overline \psi'} (p)d \tilde{\psi'} (p)=K{\cal{D}}\tilde{\overline \psi'} {\cal{D}}\tilde{\psi'}$.} Therefore, it follows from the above discussion that \eqref{Gamma_star} can be simply written as
\begin{equation}
e^{i\Gamma_{\star'}[A]}=\frac{{\int {\cal{D}}\overline \psi{\cal{D}}\psi ~e^{i{\cal{S}}_{f,\star'}(\overline \psi,\psi,A)}}}{{\int {\cal{D}}\overline \psi {\cal{D}}\psi ~e^{i{\cal{S}}_{f,\star'}(\overline \psi,\psi,0)}}}=\frac{{\int {\cal{D}}\overline \psi'{\cal{D}}\psi' ~e^{i{\cal{S}}_{f,\star}(\overline \psi',\psi',A')}}}{{\int {\cal{D}}\overline \psi' {\cal{D}}\psi' ~e^{i{\cal{S}}_{f,\star}(\overline \psi',\psi',0)}}}=e^{i\Gamma_{\star}[A']},
\label{Gamma_star-2}
\end{equation}
or more precisely we find that $\Gamma_{\star'}[A]=\Gamma_{\star}[A']$, a result that naturally leads to
 \begin{equation} \label{S_star'=S_star}
S_{\star'}=S_{\star}.
\end{equation}
We then see that that the result, Eq.~\eqref{S_star'=S_star}, holds as long the above statements are true.

%%%%%%%%%%%%%%%%%%%%%%%%%%%%%%%%%%%%%%%%%%%%%%%%%%%%%%%%%%%
\section{Final remarks}
\label{sec5}

In this paper, we have discussed the noncommutative QED$_2$ in a S-matrix framework.
This powerful tool was used in order to establish a proper analysis on the dynamical mass generation for the gauge field, where we were interested in determining the exactness of Schwinger mass $\mu^2=\frac{e^2}{\pi}$ and that it does not receive noncommutative corrections at any loop order. In this sense the S-matrix approach was rather helpful since it allows us to work with the effective action $\Gamma[A]$ (interaction term), and that S-matrix elements correspond to the desired contributions (graphs) to the 1PI gauge function.

Our main interest was twofold: first, we wanted to establish that the Schwinger mass is perturbatively exact in noncommutative spacetime (as well in the commutative one, see appendix \ref{sec-appA}); second, we wished to use the computation of S-matrix elements to fully determine the noncommutative phase factors, which in general are quite complicated to obtain when Feynman rules are used. In this sense, we have divided our analysis in two parts: one-fermionic loop and multi-fermionic loop, because in this way we are able to highlight all the necessary aspects in order to show that  1PI gauge functions are vanishing at higher loops, and only a planar graph is present at one-loop order contributing to the generation of the Schwinger mass.

 Because of the importance of the obtained results to the noncommutative Schwinger mass, we used $\alpha^*$-cohomology results to generalize our analysis on the Moyal star product to all translation-invariant star products. This approach allowed us to show that $S_{\star'}=S_{\star}$, i.e., that all S-matrix elements are equal in a class of theories with translation-invariant star products.
 \newpage
%%%%%%%%%%%%%%%%%%%%%%%%%%%%%%%%%%%%%%%%%%%%%%%%%%%%%%%%%%%
\subsection*{Acknowledgments}
The authors are thankful to M. Chaichian for his support and interest in this work as well as his comments on the
manuscript. M. Gh. is grateful to M. M. Sheikh-Jabbari for fruitful discussion. Moreover, encouraging comments
from C. P. Martin are kindly appreciated. R.B. acknowledges partial support from Conselho Nacional de Desenvolvimento Cient\'ifico e Tecnol\'ogico (CNPq Project No. 304241/2016-4) and Funda\c{c}\~ao de Amparo \`a Pesquisa do Estado de Minas Gerais (FAPEMIG Project No. APQ-01142-17).
%%%%%%%%%%%%%%%%%%%%%%%%%%%%%%%%%%%%%%%%%%%%%%%%%%%%%%%%%%

\appendix

\section{A perturbative proof on the exactness of the Schwinger mass}
\label{sec-appA}

In 1962, Schwinger showed that, when the spacetime dimensionality is lowered, the gauge invariance does not necessarily impose a zero mass for the gauge particle \cite{schwinger-1,schwinger-2}. He found an exact value for the photon mass in $1+1$ dimensions by making use of a nonperturbative method.

On the perturbative side, the one-loop computation of the Schwinger mass can be found in \cite{peskin} but the higher-order perturbative proof on the exactness of the Schwinger mass has not yet been studied.\\
\noindent
In this appendix, we present a technical method, as a result of the super-renormalizability of QED$_{2}$, and perturbatively show that the higher-order contributions to the photon self-energy and the Schwinger mass are vanished. Therefore, the Schwinger mass is one-loop exact and its value is the same as the result found by the nonperturbative analysis in \cite{schwinger-2}.\\
\noindent
Let us start by defining the Lagrangian density of the massless commutative $QED_{2}$,
\begin{equation}
{\cal L}=i\bar\psi\gamma^{\mu}\partial_{\mu}\psi-e\bar\psi\gamma^{\mu}A_{\mu}\psi-\frac{1}{4}F_{\mu\nu}F^{\mu\nu},
\label{eq:c1}
\end{equation}
where $\gamma^{\mu}$'s are two-dimensional matrices
\begin{equation}
\gamma^{0}=\left(
  \begin{array}{cc}
    0 & 1 \\
   1 & 0\\
  \end{array}
\right),~~~~~~ \gamma^{1}=\left(
  \begin{array}{cc}
    0 & -1 \\
    1 & 0\\
  \end{array}
\right).
\label{eq:c2}
\end{equation}
The general structure of the photon self-energy is given as usual
\begin{equation}
\Pi_{\mu\nu}(q^{2})= \left(q^{2}g_{\mu\nu}-q_{\mu}q_{\nu}\right)\Pi(q^{2}).
\label{eq:c3}
\end{equation}
The one-loop calculation yields us $\Pi^{(1)} (q)=\frac{e^{2}}{\pi q^{2}}$ and this generates a nonzero pole for the one-loop corrected propagator written as
\begin{equation}
{\cal {D}}_{\mu\nu}^{(1)}(q)=\frac{-ig_{\mu\nu}}{q^{2}-\frac{e^{2}}{\pi}}.
\label{eq:c4}
\end{equation}
For comparison, the produced pole in the denominator of the one-loop propagator is exactly the Schwinager mass and hence we expect that higher-loop contributions to this mass vanish exactly.
Hence, we shall discuss next the two- and three-loop orders,
and similarly show that the next orders not only do not correct the mass but are also identically zero.

%%%%%%%%%%%%%%%%%%%%%%%%%%%%%%%%%%%%%%%%
\subsection{Two-loop contribution}
%%%%%%%%%%%%%%%%%%%%%%%%%%%%%%%%%%%%%%%%

\begin{figure}[t]
 \centering
  \includegraphics[width=11.2cm,height=2.4cm]{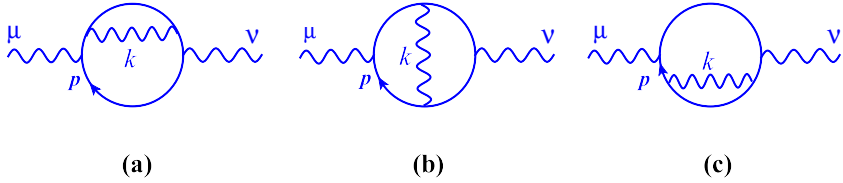}
    \caption{The graphs (a), (b), and (c) represent the two-loop contributions
to the photon self-energy.
}\label{2-loop}
\end{figure}

Let us consider the two-loop graphs, depicted in Fig. \ref{2-loop}. The total two-loop contribution is given by
\begin{equation}
\Pi_{\mu\nu}^{(2)}(q)=\sum_{i=a,b,c}\Pi_{\mu\nu}^{(2,i)}(q),
\label{eq:c5}
\end{equation}
where the amplitudes are given by
\begin{align}
\Pi_{\mu\nu}^{(2,a)}(q)&=e^{4}\int\frac{d^{2}p}{(2\pi)^{2}}\frac{d^{2}k}{(2\pi)^{2}}
~\frac{{\cal{N}}_{\mu\nu}^{\tiny\mbox{(a)}}}{k^{2}(q+p)^{2}(q+p+k)^{2}(q+p)^{2}p^{2}},
\nonumber\\
\Pi_{\mu\nu}^{(2,b)}(q)&=e^{4}\int\frac{d^{2}p}{(2\pi)^{2}}\frac{d^{2}k}{(2\pi)^{2}}
~\frac{{\cal{N}}_{\mu\nu}^{\tiny\mbox{(b)}}}{k^{2}(q+p)^{2}(q+p+k)^{2}(p+k)^{2}p^{2}},
\nonumber\\
\Pi_{\mu\nu}^{(2,c)}(q)&=e^{4}\int\frac{d^{2}p}{(2\pi)^{2}}\frac{d^{2}k}{(2\pi)^{2}}
~\frac{{\cal{N}}_{\mu\nu}^{\tiny\mbox{(c)}}}{k^{2}(q+p)^{2}(p+k)^{2}p^{4}},
\label{eq:c6}
\end{align}
with the simplified notation
\begin{align}
{\cal{N}}_{\mu\nu}^{(a)}&=tr\Big(\gamma_{\mu}
(\displaystyle{\not}q+\displaystyle{\not}p)
\gamma^{\rho}(\displaystyle{\not}q+\displaystyle{\not}p+\displaystyle{\not}k)
\gamma_{\rho}(\displaystyle{\not}q+\displaystyle{\not}p)\gamma_{\nu}
\displaystyle{\not}p\Big),
\nonumber\\
%%%
{\cal{N}}_{\mu\nu}^{(b)}&=tr\Big(\gamma_{\mu}
(\displaystyle{\not}q+\displaystyle{\not}p)
\gamma^{\rho}(\displaystyle{\not}q+\displaystyle{\not}p+\displaystyle{\not}k)
\gamma_{\nu}(\displaystyle{\not}p+\displaystyle{\not}k)\gamma_{\rho}
\displaystyle{\not}p\Big),
\nonumber\\
%%%
{\cal{N}}_{\mu\nu}^{(c)}&=tr\Big(\gamma_{\mu}
(\displaystyle{\not}q+\displaystyle{\not}p)
\gamma_{\nu}\displaystyle{\not}p\gamma^{\rho}(\displaystyle{\not}p+\displaystyle{\not}k)
\displaystyle{\not}p\gamma_{\rho}
\Big).
\label{eq:c7}
\end{align}
%%%%%%%%%%%%%%%%%%%%%%%%%%%%%%
%%%%%%%%%%%%%%%%%%%%%%%%%%%%%%

Since $QED_{2}$ is a super-renormalizable field theory, we can use the explicit form of the gamma matrices to simplify significantly the higher-loop contributions.
Thus, we need to find a covariant form for $\gamma^{\mu}\displaystyle{\not}p$ to insert back into Eq.~\eqref{eq:c6}. Thus, at the first step we can write
 \begin{equation}
\displaystyle{\not}p=p_{0}\gamma^{0}+p_{1}\gamma^{1} = \left(
  \begin{array}{cc}
    0 & p_{-} \\
    p_{+} & 0\\
  \end{array}
\right),
\label{eq:c8}
\end{equation}
where, $p_{\pm}=p_{0}\pm p_{1}$. Multiplying $\displaystyle{\not}p$ by $\gamma^{0}$ and $\gamma^{1}$, we have
\begin{equation}
\gamma^{0}\displaystyle{\not}p= \left(
  \begin{array}{cc}
     p_{+}& 0 \\
     0& p_{-}\\
  \end{array}
\right),\quad
\gamma^{1}\displaystyle{\not}p= \left(
  \begin{array}{cc}
     -p_{+}& 0 \\
     0& p_{-}\\
  \end{array}
\right).
\label{eq:c9}
\end{equation}
Now, we introduce a convenient set of light-like vectors defined as\footnote{Here, we thank to F. Ardalan for introducing this notation.}
\begin{equation}
u^{0}=1,\quad\bar u^{0}=1,\quad u^{1}=1,\quad \bar u^{1}=-1,
\label{eq:c10}
\end{equation}
and allow us to finally arrive at
\begin{equation}
\gamma^{\mu}\displaystyle{\not}p= \left(
  \begin{array}{cc}
     \bar u^{\mu}p_{+}& 0 \\
     0& u^{\mu}p_{-}\\
  \end{array}
\right),
\label{eq:c11}
\end{equation}
which is a diagonal matrix. Equivalently, we define
\begin{equation}
u_{0}=1,\quad \bar u_{0}=1,\quad u_{1}=-1,\quad \bar u_{1}=1,
\label{eq:c12}
\end{equation}
and we have
\begin{equation}
\gamma_{\mu}\displaystyle{\not}p= \left(
  \begin{array}{cc}
     \bar u_{\mu}p_{+}& 0 \\
     0& u_{\mu}p_{-}\\
  \end{array}
\right).
\label{eq:c13}
\end{equation}
Furthermore, it is easy to verify that
\begin{equation}
u_{\mu}u^{\mu}=0,\quad \bar u_{\mu}\bar u^{\mu}=0.
\label{eq:c13.5}
\end{equation}
 Actually, the light-like property of $u$ and $\bar u$ helps us to simplify significantly our loop calculations. These light-like vectors permit us to rewrite the numerators of Eq.~\eqref{eq:c7} simply in terms of
\begin{equation}
{\cal{N}}_{\mu\nu}^{(a)}=tr
\left(
  \begin{array}{cc}
       a_{1}\bar\lambda_{\mu\nu}& 0 \\
     0& a_{2}\lambda_{\mu\nu}\\
  \end{array}
\right),\quad
{\cal{N}}_{\mu\nu}^{(b)}=tr
\left(
  \begin{array}{cc}
      b_{1}\bar\lambda_{\mu\nu}& 0 \\
     0& b_{2}\lambda_{\mu\nu}\\
  \end{array}
\right),\quad
{\cal{N}}_{\mu\nu}^{(c)}=tr
\left(
  \begin{array}{cc}
      c_{1}\bar\lambda_{\mu\nu}& 0 \\
     0& c_{2}\lambda_{\mu\nu}\\
  \end{array}
\right),
\label{eq:c14}
\end{equation}
 where we have introduced the following notation: $\bar\lambda_{\mu\nu}=\bar u_{\mu} \bar u_{\nu}\bar u^{\rho}\bar u_{\rho}$ and $\lambda_{\mu\nu}=u_{\mu}u_{\nu}u^{\rho}u_{\rho}$ and, conveniently, the coefficients $a_i$, $b_i$, and $c_i$, with $i=1,2$, which are given by
\begin{align}
&\left\{
  \begin{array}{llll}
    a_{1}=(q+p)_{+} (q+p+k)_{+}(q+p)_{+}p_{+},\quad
    a_{2}=(q+p)_{-} (q+p+k)_{-}(q+p)_{-}p_{-},\\
    b_{1}= (q+p)_{+} (q+p+k)_{+}(p+k)_{+}p_{+},\quad
    b_{2}=(q+p)_{-} (q+p+k)_{-}(p+k)_{-}p_{-},\\
     c_{1}=(q+p)_{+} p_{+}(p+k)_{+}p_{+},\quad\quad\quad\quad\quad~
    c_{2}=(q+p)_{-} p_{-}(p+k)_{-}p_{-}.\\
  \end{array}
\right.
\label{eq:c15}
\end{align}
We then see that with the assistance of the above light-like vectors and the identity equation \eqref{eq:c13}, the complicated form of the three numerators appearing in Eq.~\eqref{eq:c7} can be put into a simple diagonal form, as seen in the revised expressions in Eq.~\eqref{eq:c14}.

Finally, according to Eq.~\eqref{eq:c13.5}, it is easy to realize that $\bar\lambda_{\mu\nu}=\lambda_{\mu\nu}=0$, which lead to
${\cal{N}}_{\mu\nu}^{(a)}={\cal{N}}_{\mu\nu}^{(b)}={\cal{N}}_{\mu\nu}^{(c)}=0$. Hence, without performing any integration, just a simple $\gamma$-matrix algebra, we obtain
\begin{equation}
\Pi_{\mu\nu}^{(2,a)}(q)=\Pi_{\mu\nu}^{(2,b)}(q)=\Pi_{\mu\nu}^{(2,c)}(q)=0.
\label{eq:c16}
\end{equation}
Therefore, we can conclude that $\Pi_{\mu\nu}^{(2)}(q)=0$. This result indicates that the Schwinger mass does not receive corrections at two-loop level. In what follows, we study the three-loop analysis contributing to the photon self-energy sector.

%%%%%%%%%%%%%%%%%%%%%%%%%%%%%
\subsection{Three-loop contribution}
%%%%%%%%%%%%%%%%%%%%%%%%%%%%%

\begin{figure}[t]
 \centering\vspace{-1cm}
  \includegraphics[width=11cm,height=4.5cm]{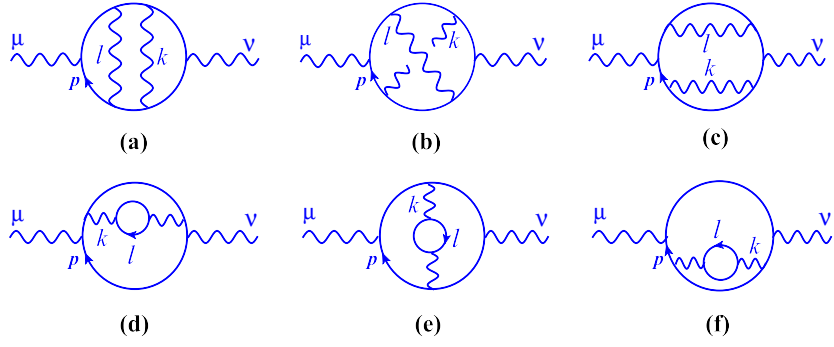}
    \caption{The graphs (a)-(f) represent the one and two fermion-loop contributions to the photon self-energy at three-loop level.
}\label{3-loop}
\end{figure}

The relevant set of graphs for the three-loop contribution for the photon self-energy are shown in Fig.~\ref{3-loop}. The Feynman expressions for these graphs are given by
\begin{equation}
\Pi_{\mu\nu}^{(3)}=\sum_{i=a,b,c}\Pi_{\mu\nu}^{(3,i)}+\sum_{i=d,e,f}\Pi_{\mu\nu}^{(3,i)}.
\label{eq:c17}
\end{equation}
The first and the second sums refer to the diagrams with one and two fermion loops (remember from the single- and multi-fermionic loop discussion), respectively. Now, it is easily found that the first sum is given by the expression
%%%%%%%%%%%%%%%%%%%%%%%%%%%%%%
%%%%%%%%%%%%%%%%%%%%%%%%%%%%%%
%%%%%%%%%%%%%%%%%%%%%%%%%%%%%%%%%%%%%%%%%%%%%%%%%%%%%%%%%%%%%%%%%%%%%%%%%%%%%%%%%%
\begin{align}
\Pi_{\mu\nu}^{(3,a+b+c)}(q)&=e^{6}\int\frac{d^{2}p}{(2\pi)^{2}}\frac{d^{2}k}{(2\pi)^{2}}\frac{d^{2}\ell}{(2\pi)^{2}}
~\frac{1}
{k^{2}\ell^{2}p^{2}(q+p)^{2}(q+p+\ell)^{2}}\nonumber\\
&\times
\Bigg\{\frac{1}{(q+p+\ell+k)^{2}(p+\ell+k)^{2}}\Big[\frac{{\cal N}_{\mu\nu}^{(a)}}{(p+\ell)^{2}}+\frac{{\cal N}_{\mu\nu}^{(b)}}{(p+k)^{2}}\Big]+
\frac{{\cal N}_{\mu\nu}^{(c)}}
{(p+q)^{2}(p+k)^{2}p^{2}}
\Bigg\},
\label{eq:c18}
\end{align}
with the definition
\begin{align}
{\cal N}_{\mu\nu}^{(a)}&=tr\Big(\gamma_{\mu}
(\displaystyle{\not}q+\displaystyle{\not}p)\gamma^{\rho}(\displaystyle{\not}q+\displaystyle{\not}p+
\displaystyle{\not}\ell)\gamma^{\sigma}(\displaystyle{\not}q+\displaystyle{\not}p+\displaystyle{\not}\ell+\displaystyle{\not}k)
\gamma_{\nu}(\displaystyle{\not}p+\displaystyle{\not}\ell+\displaystyle{\not}k)\gamma_{\sigma}(\displaystyle{\not}p+
\displaystyle{\not}\ell)\gamma_{\rho}
\displaystyle{\not}p\Big),
\nonumber\\
{\cal N}_{\mu\nu}^{(b)}&=tr\Big(\gamma_{\mu}
(\displaystyle{\not}p+\displaystyle{\not}q)\gamma^{\rho}
(\displaystyle{\not}p+\displaystyle{\not}q+\displaystyle{\not}\ell)\gamma^{\sigma}
(\displaystyle{\not}p+\displaystyle{\not}q+\displaystyle{\not}\ell+\displaystyle{\not}k)
\gamma_{\nu}(\displaystyle{\not}p+\displaystyle{\not}\ell+\displaystyle{\not}k
)
\gamma_{\rho}(\displaystyle{\not}p+\displaystyle{\not}k)\gamma_{\sigma}\displaystyle{\not}p\Big),
\nonumber\\
{\cal N}_{\mu\nu}^{(c)}&=tr\Big(\gamma_{\mu}
(\displaystyle{\not}p+\displaystyle{\not}q)
\gamma_{\rho}(\displaystyle{\not}q+\displaystyle{\not}p+\displaystyle{\not}\ell)
\gamma^{\rho}(\displaystyle{\not}p+\displaystyle{\not}q)\gamma_{\nu}
\displaystyle{\not}p\gamma^{\sigma}(\displaystyle{\not}p+\displaystyle{\not}k)\gamma_{\sigma}\displaystyle{\not}p\Big).
\label{eq:c19}
\end{align}
Also, the second sum, the two-fermionic-loop part, is presented as
\begin{equation}
\Pi_{\mu\nu}^{(3,d+e+f)}(q)=
e^{6}\int\frac{d^{2}p}{(2\pi)^{2}}\frac{d^{2}k}{(2\pi)^{2}}
\frac{{\cal D}^{\rho\sigma}_{\tiny\mbox{(1)}}(k)}
{p^{2}(p+q)^{2}}
\Bigg\{\frac{1}{(q+p+k)^{2}}
\left[
\frac{{\cal N}_{\mu\nu\rho\sigma}^{(d)}}
{(p+q)^{2}}
+\frac{{\cal N}_{\mu\nu\rho\sigma}^{(e)}}
{(p+k)^{2}}\right]+
\frac{{\cal N}_{\mu\nu\rho\sigma}^{(f)}}
{p^{2}(p+k)^{2}}
\Bigg\},
\label{eq:c20}
\end{equation}
where $i{\cal D}^{\rho\sigma}_{(1)}(k)=\frac{g^{\rho\sigma}}{k^{2}-\frac{e^{2}}{\pi}}$ is representing the one-loop corrected photon propagator, as mentioned in Eq.~\eqref{eq:c4}. Thus, we can rewrite Eq.~\eqref{eq:c20} explicitly as
\begin{align}
\Pi_{\mu\nu}^{(3,d+e+f)}(q)&=
-ie^{6}\int\frac{d^{2}p}{(2\pi)^{2}}\frac{d^{2}k}{(2\pi)^{2}}~~
\frac{1}
{p^{2}(p+q)^{2}(k^{2}-\frac{e^{2}}{\pi})}
\nonumber\\
&\times\Bigg\{\frac{1}{(q+p+k)^{2}}
\Big[
\frac{\widetilde{{\cal N}}_{\mu\nu}^{(d)}}
{(p+q)^{2}}
+\frac{\widetilde{{\cal N}}_{\mu\nu}^{(e)}}
{(p+k)^{2}}\Big]+
\frac{\widetilde{{\cal N}}_{\mu\nu}^{(f)}}
{p^{2}(p+k)^{2}}
\Bigg\},
\label{eq:c21}
\end{align}
where we have defined $\widetilde{{\cal N}}_{\mu\nu}=g^{\rho\sigma}{\cal N}_{\mu\nu\rho\sigma}$, which read explicitly
\begin{align}
\widetilde{{\cal N}}_{\mu\nu}^{(d)}&=tr\Big(\gamma_{\mu}
(\displaystyle{\not}p+\displaystyle{\not}q)\gamma^{\rho}(\displaystyle{\not}p+\displaystyle{\not}q+
\displaystyle{\not}k)\gamma_{\rho}(\displaystyle{\not}p+\displaystyle{\not}q)
\gamma_{\nu}\displaystyle{\not}p\Big),\nonumber\\
\widetilde{{\cal N}}_{\mu\nu}^{(e)}&=tr\Big(\gamma_{\mu}
(\displaystyle{\not}p+\displaystyle{\not}q)\gamma^{\rho}(\displaystyle{\not}p+\displaystyle{\not}q+
\displaystyle{\not}k)\gamma_{\nu}(\displaystyle{\not}p+\displaystyle{\not}k)
\gamma_{\rho}\displaystyle{\not}p\Big),\nonumber\\
\widetilde{{\cal N}}_{\mu\nu}^{(f)}&=tr\Big(\gamma_{\mu}
(\displaystyle{\not}p+\displaystyle{\not}q)\gamma_{\nu}\displaystyle{\not}p
\gamma^{\rho}(\displaystyle{\not}p+\displaystyle{\not}k)
\gamma_{\rho}\displaystyle{\not}p\Big).
\label{eq:c22}
\end{align}
The general structure of the revised numerators in terms of the light-like vectors $u$ and $\bar u$ is classified into two cases. For the Eq.~\eqref{eq:c19}, it is easy to verify that
\begin{equation}
{\cal{N}}_{\mu\nu}^{(j)}=tr
\left(
  \begin{array}{cc}
       \alpha_{1}^{(j)}\bar\beta_{\mu\nu}& 0 \\
     0& \alpha_{2}^{(j)}\beta_{\mu\nu}\\
  \end{array}
\right),\quad j=a,b,c
\label{eq:c23}
\end{equation}
with $\bar\beta_{\mu\nu}=\bar u_{\mu}\bar u_{\nu}\bar u^{\rho}\bar u_{\rho}\bar u^{\sigma}\bar u_{\sigma}$ and $\beta_{\mu\nu}=u_{\mu}u_{\nu}u^{\rho}u_{\rho}u^{\sigma}u_{\sigma}$. Now, proceeding in the same lines as before, for the expressions in Eq.~\eqref{eq:c22}, we get
%%%
\begin{equation}
\widetilde{{\cal{N}}}_{\mu\nu}^{(j)}=tr
\left(
  \begin{array}{cc}
       \eta_{1}^{(j)}\bar\omega_{\mu\nu}& 0 \\
     0& \eta_{2}^{(j)}\omega_{\mu\nu}\\
  \end{array}
\right),\quad j=d,e,f
\label{eq:c24}
\end{equation}
with $\bar\omega_{\mu\nu}=\bar u_{\mu}\bar u_{\nu}\bar u^{\rho}\bar u_{\rho}$ and $\omega_{\mu\nu}=u_{\mu}u_{\nu}u^{\rho}u_{\rho}$. The quantities $\alpha_{1}^{(j)}$, $\alpha_{2}^{(j)}$, $\eta_{1}^{(j)}$, and
$\eta_{2}^{(j)}$ are defined as a function of the momenta $(p,q,k,\ell)$.

Besides, due to Eq.~\eqref{eq:c13.5}, it is easily observed that $\bar\beta_{\mu\nu}=\beta_{\mu\nu}=0$ and $\bar\omega_{\mu\nu}=\omega_{\mu\nu}=0$. This result leads to the result $\Pi_{\mu\nu}^{(3)}=0$ and again we observe that no  quantum correction to the Schwinger mass is present at this order too. Next, we discuss our analysis for the higher-loop order.

%%%%%%%%%%%%%%%%%%%%%%%%%%%%%
\subsection{Higher-loop contribution}
%%%%%%%%%%%%%%%%%%%%%%%%%%%%%
%%%%%%%%%%%%%%%%%%%%%%%%%%%%%%
\begin{figure}[t]
 \centering\vspace{-1cm}
  \includegraphics[width=5.2cm,height=3cm]{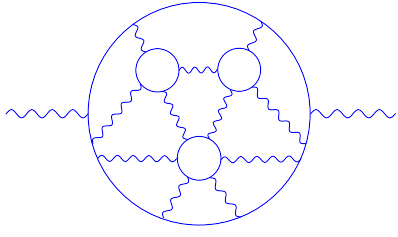}
    \caption{Higher-loop contribution to the photon self-energy.
}\label{higher-loop}
\end{figure}
%%%%%%%%%%%%%%%%%%%%%%%%%%%%%%%%%%%%%%%%%%
After the previous discussions we are ready to study a general diagram with an arbitrary number of loops, contributing to the photon self-energy. For example, let us consider a complex graph of an arbitrary order $n$, depicted in Fig.~\ref{higher-loop}, and write qualitatively its Feynman expression
\begin{equation}
\Pi_{\mu\nu}^{(n)}(q)\propto \int\prod_{i}\frac{d^{2}k_{i}}{(2\pi)^{2}}~\frac{{\cal N}_{\mu\nu}(q,k_{i})}{{\cal M}(q,k_{i})},
\label{eq:c25}
\end{equation}
where ${\cal N}_{\mu\nu}(q,k_{i})=tr\Big(\gamma_{\mu}\displaystyle{\not}{\cal B}_{1}\gamma^{\rho}\displaystyle{\not}{\cal B}_{2}\gamma^{\sigma}\displaystyle{\not}{\cal B}_{3}\gamma_{\nu}\ldots \gamma_{\rho}\displaystyle{\not}{\cal B}_{m}\gamma_{\sigma} \Big)$ and $\displaystyle{\not}{\cal B}_{i}=\displaystyle{\not}{\cal B}_{i}(q,k_{i})$.
We notice that the presence of one photon propagator at least, which is proportional to the metric, forces a contraction of the indices giving either $\bar u^{\rho}\bar u_{\rho}=0$, or $u^{\rho}u_{\rho}=0$ and consequently resulting into $\Pi_{\mu\nu}^{(n)}(q)=0$.

Indeed, this is a key point of our proposed method that yields results without carrying out any integration over the internal loop momenta.
As an important result of our present perturbative analysis, we conclude that the vacuum polarization tensor in two dimensions receives quantum corrections only from one-loop level. Hence the photon propagator in Eq.~\eqref{eq:c4} is indeed one-loop exact and, accordingly, its nonzero pole is also exact. This result is in agreement
with the Schwinger result \cite{schwinger-2}, obtained by a nonperturbative analysis.

%%%%%%%%%%%%%%%%%%%%%%%%%%%%%%%%%%%%%%%%%%%%%%%%%%%%%%%%%%
\section{Detailed analysis of the integrals over the fermion loop}
\label{sec-appB}

In this appendix, we intend to carefully analyze the integrals over the fermion loop appearing in our loop calculation and prove that they actually vanish. For simplicity, we choose to illustrate the analysis of the second equation of \eqref{eq:b17}, which was also discussed in \cite{ghasemkhani-13},
 \begin{equation}
\widetilde{\Xi}_{\mu\rho}^{(b)}=\int\frac{d^{2}p}{(2\pi)^{2}}~\frac{tr\Big(\gamma^{\mu}
(\displaystyle{\not}p+\displaystyle{\not}q)
\gamma^{\nu}(\displaystyle{\not}p+\displaystyle{\not}q+\displaystyle{\not}\ell)
\gamma^{\rho}(\displaystyle{\not}p+\displaystyle{\not}\ell)
\gamma_{\nu}\displaystyle{\not}p\Big)
}{(p+q)^{2}(p+q+\ell)^{2}(p+\ell)^{2}p^{2}}.
\end{equation}
According to the aforementioned discussion in Ref.~\cite{ghasemkhani-13}, we first revise this integral in the light-cone coordinate as
\begin{equation}
\widetilde{\Xi}_{--}^{(b)}=\int\frac{d^{2}p}{(2\pi)^{2}}~\frac{tr\Big(\gamma_{-}
(\displaystyle{\not}p+\displaystyle{\not}q)
\gamma_{-}(\displaystyle{\not}p+\displaystyle{\not}q+\displaystyle{\not}\ell)
\gamma_{-}(\displaystyle{\not}p+\displaystyle{\not}\ell)
\gamma_{-}\displaystyle{\not}p\Big)
}{(p+q)^{2}(p+q+\ell)^{2}(p+\ell)^{2}p^{2}}.
\end{equation}
Using $p^{2}=p_{+}p_{-}$ and the related details mentioned in appendix A of \cite{ghasemkhani-13}, we arrive at the expression
\begin{equation}
\widetilde{\Xi}_{--}^{(b)}=16\int\frac{dp_{+}}{(2\pi)}\frac{dp_{-}}{(2\pi)}~
\frac{1}{(p+q)_{+}(p+q+\ell)_{+}(p+\ell)_{+}p_{+}}.
\end{equation}
In order to simplify this result, we decompose the fraction into partial fractions to reduce the degree of the denominator. Thus, we find
\begin{align}
\widetilde{\Xi}_{--}^{(b)}&=16\int\frac{dp_{-}}{2\pi}\frac{
dp_{+}}{2\pi}~\frac{1}{\ell^{2}_{+}} ~\bigg\{
\frac{1}{q_{+}}\left(\frac{1}{p_{+}}-\frac{1}{(p+q)_{+}}\right)
-\frac{1}{(\ell-q)_{+}}\left(\frac{1}{(p+q)_{+}}-\frac{1}{(p+\ell)_{+}}\right)
\nonumber\\
&-\frac{1}{(\ell+q)_{+}}\left(\frac{1}{p_{+}}-\frac{1}{(p+q+\ell)_{+}}\right)
+\frac{1}{q_{+}}\left(\frac{1}{(p+\ell)_{+}}-\frac{1}{(p+q+\ell)_{+}}\right)\bigg\}.\label{eq:d1a}
\end{align}

\begin{figure}[t]
\vspace{-1.4cm}
\hspace{-2cm}
\centering\includegraphics[width=5cm,height=3.5cm]{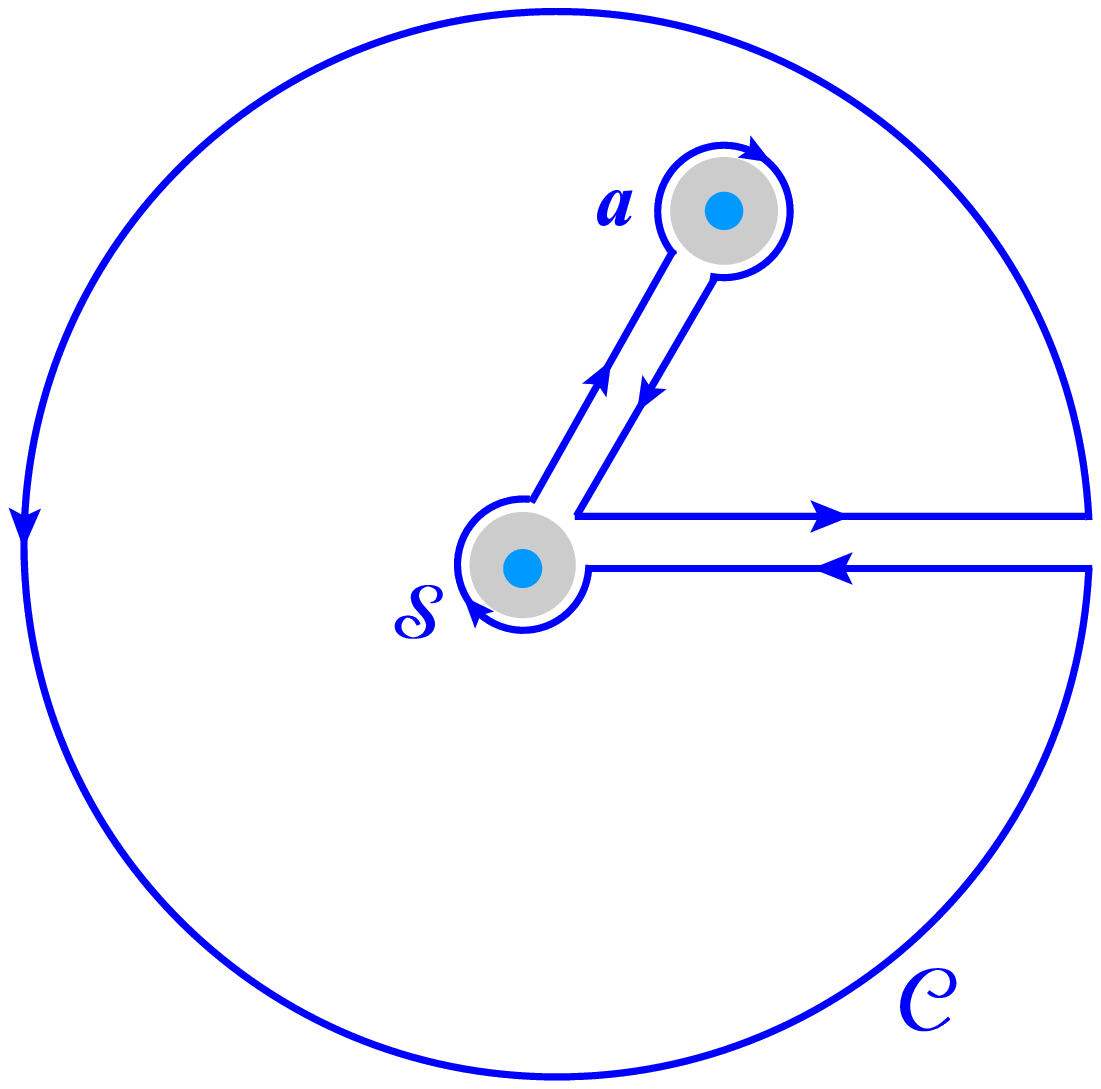}
      \caption{The relevant contour for the integral \eqref{eq:d2}.}
      \label{circle}
\end{figure}

Now, we will show that each separated pairs in the parentheses are identically zero and we are left with $\widetilde{\Xi}_{--}^{(b)}=0$.
 To prove this result in the simplest way we recall a technical theorem of standard calculus named as \emph{complex form of Green's theorem}, which is illustrated below \cite{spiegel}:

Suppose that $B(z,\bar{z}):\mathbb{C}\rightarrow\mathbb{C}$ is continuous and has continuous partial derivatives in a region $\mathcal{R}\subsetneq
\mathbb{C}$ and on its boundary $\mathcal{C}$, then:
\begin{equation}
\oint_{\mathcal{C}} B(z,\bar{z})dz=2i\iint_{\mathcal{R}}
\frac{\partial B}{\partial \bar{z}}dxdy.
\label{eq:d1}
\end{equation}
Now let us assume that the function $B(z,\bar{z})$ coincides with
$\bar{z}\left(\frac{1}{z}-\frac{1}{z-a}\right)$  over the region $\mathcal{R}$,
as shown in Fig.~\ref{circle}, for some given
$a~\epsilon~\mathbb{C}$ . As it may be clear, the region is
essentially a large disc with radius $r\rightarrow\infty$, with which two
specific points, the origin and the simple pole $a~\epsilon~\mathbb{C}$, as two smaller
discs, have been punctured out from (the subsets which have been colored
by gray). Let us show the rest part of the disc, the union of gray
colored points, with $\mathcal{S}$, as shown in Fig.~\ref{circle}.

Now by the above theorem we can conclude that
\begin{equation}
\iint_{\mathcal{R}}\left(\frac{1}{z-a}-\frac{1}{z}\right)~dxdy=
\frac{i}{2}\oint_{\mathcal{C}}\bar{z}\left(\frac{1}{z}-\frac{1}{z-a}\right)dz.
\label{eq:d2}
\end{equation}
Here we should suppose that the area of $\mathcal{S}$ is tiny
enough, i.e. the radiuses of gray discs, commonly shown by
$\varepsilon$, tend to zero. Therefore, it is also easy to show that
\begin{equation}
\lim_{\varepsilon\rightarrow
0}\iint_{\mathcal{S}}\left(\frac{1}{z-a}-\frac{1}{z}\right)~dxdy=0.
\label{eq:d3}
\end{equation}
%%%%%%%%%%%%%%%%%%%%%%%%%%%%%%%%%%%%%%%%%%%%%%%%%%%%%%%%%%%%%%%%
%%%%%%%%%%%%%%%%%%%%%%%%%%%%%%%%%%%%%%%%%%%%%%%%%%%%%%%%%%%%%%%%
In fact, if ${\cal{S}}_{0}$ (resp. ${\cal{S}}_{a}$) is the
punctured disc at the origin (resp. $a$), for $\frac{\mid
a\mid}{2}\geq\varepsilon$  we have
\begin{align}
\big\vert\iint_{{\cal S}}\bigg(\frac{1}{re^{i\theta}-a}-\frac{1}{re^{i\theta}}\bigg)rdrd\theta\big\vert
&\leq \iint_{\cal S}\big\vert\bigg(\frac{1}{re^{i\theta}-a}-\frac{1}{re^{i\theta}}\bigg)\big\vert ~ rdrd\theta\nonumber\\
&\leq \int_{0}^{2\pi}
d\theta\int_{0}^{\varepsilon}\bigg(\frac{2}{a}+\frac{1}{r}\bigg)rdr\nonumber\\
&=2\pi \varepsilon \left(\frac{\varepsilon}{a}+1\right),
\label{eq:d4}
\end{align}
which clearly vanishes as $\varepsilon\rightarrow 0$. The same
assertion can be simply stated for the integration over
$\mathcal{S}_{a}$. Therefore, Eq.~\eqref{eq:d2} is modified to
\begin{equation}
\iint_{\mathbb{R}^{2}=\mathbb{C}}\left(\frac{1}{z-a}-\frac{1}{z}\right)~dxdy=\frac{i}{2}\lim_{r\rightarrow
\infty}\lim_{\varepsilon\rightarrow 0}
\oint_{\mathcal{C}}\bar{z}\left(\frac{1}{z}-\frac{1}{z-a}\right)~dz,
\label{eq:d5}
\end{equation}
where the lhs is exactly what we need in our manipulations. It is
enough to calculate precisely the rhs for $\mathcal{C}$. Let us
divide $\mathcal{C}$ to three different circles: (a)
$\mathcal{C}_{0}$, the boundary of $\mathcal{S}_{0}$, (b)
$\mathcal{C}_{a}$, the boundary of $\mathcal{S}_{a}$, and (c)
$\mathcal{C}'$ the large circle with radius $r\rightarrow\infty$. As
above, according to the symmetry, the solution of part (a) works
similarly for (b). Therefore, we should work out (a) and (c). For
part (a) we have that
\begin{align}
\lim_{\varepsilon\rightarrow
0}\big\vert\oint_{\mathcal{C}_{0}}\bar{z}\left(\frac{1}{z}-\frac{1}{z-a}\right)~dz\big\vert&=i\lim_{\varepsilon\rightarrow
0}
\big\vert\int_{0}^{2\pi}\varepsilon^{2}\bigg(\frac{e^{-i\theta}}{\varepsilon}-\frac{1}{\varepsilon
e^{i\theta}-a}\bigg)d\theta \big\vert\nonumber\\
&\leq i\lim_{\varepsilon\rightarrow
0}\varepsilon\bigg(1+\frac{2\varepsilon}{\vert
a\vert}\bigg)\int_{0}^{2\pi}d\theta=0.
\label{eq:d6}
\end{align}
Therefore, according to \eqref{eq:d5} we find
\begin{equation}
\iint_{\mathbb{R}^{2}=\mathbb{C}}\left(\frac{1}{z-a}-\frac{1}{z}\right)~dxdy=\frac{i}{2}\lim_{r\rightarrow
\infty} \oint_{\mathcal{C}'}\bar{z}\left(\frac{1}{z}-\frac{1}{z-a}\right)dz.
\label{eq:d7}
\end{equation}
Now, it is enough to calculate the rhs of Eq.~\eqref{eq:d7}. For
$r\gg\vert a\vert$  we read
\begin{align}
\oint_{\mathcal{C}'}\bar{z}\left(\frac{1}{z}-\frac{1}{z-a}\right)dz&=\int_{0}^{2\pi}r^{2}
\bigg(\frac{1}{re^{i\theta}}-\frac{1}{re^{i\theta}-a}\bigg)d\theta\nonumber\\
&=\int_{0}^{2\pi}re^{-i\theta}\bigg(1-\frac{1}{1-\frac{a}{r}e^{-i\theta}}\bigg)\nonumber\\
&=\int_{0}^{2\pi}re^{-i\theta}\sum\limits_{n=1}^{\infty}\left(\frac{a}{r}\right)^{n}e^{-in\theta}d\theta.
\label{eq:d8}
\end{align}
It is easily seen that for all $N\in \mathbb{N}$,
$f_N=re^{-i\theta}\sum\limits_{n=1}^N \big(\frac{a}{r}\big)^n
e^{-in\theta}$  belongs to $L^2(S^1)$, and $\vert f_N \vert\leq
r^2/(r-a) \in L^1(S^1)$. Consequently, due to the celebrated
\emph{dominated convergence theorem} one might conclude that
\begin{equation}
\int_{0}^{2\pi}re^{-i\theta}\sum\limits_{n=1}^{\infty}\left(\frac{a}{r}\right)^{n}
e^{-in\theta}d\theta=\sum\limits_{n=1}^{\infty}r\left(\frac{a}{r}\right)^{n}
\int_{0}^{2\pi}e^{-i(n+1)\theta}d\theta,
\label{eq:d9}
\end{equation}
which vanishes obviously. Therefore, summarizing the above results, we have that according to \eqref{eq:d5}, \eqref{eq:d7} and \eqref{eq:d9} we read
\begin{equation}
\iint_{\mathbb{R}^{2}=\mathbb{C}}\left(\frac{1}{z-a}-\frac{1}{z}\right)dxdy=0
\label{eq:d10}
\end{equation}
as we expected. Hence, momentum integrals that can be put into the form \eqref{eq:d1a} are vanishing due to the identity \eqref{eq:d10}. Similarly, it is easy to show that the remaining integrals in Eq.~\eqref{eq:b17}, $\widetilde{\Xi}_{\mu\nu}^{(a)}$ and $\widetilde{\Xi}_{\mu\sigma}^{(c)}$, are vanishing and that no noncommutative corrections to the Schwinger mass are found at two-loop order.

%%%%%%%%%%%%%%%%%%%%%%%%%%%%%%%%%%%%%%%%%%%%%%%%%%%%%%%%%%%%%%%%%%%%%%%%%%%%%

\end{document}